\def\kms{\,{\rm km}\,{\rm s}^{-1}}
\def\hmpc{\,{h^{-1} {\rm Mpc}}}
\documentclass[usenatbib]{mn2e}
\usepackage{epsfig}

\usepackage{amssymb}
\usepackage{amsmath}

\begin{document}

\title[Comparison between measured velocity and gravity fields]{%
A comparison of the galaxy peculiar velocity field with the PSCz
gravity field-- A Bayesian hyper-parameter method}

\author[Ma, Branchini \& Scott]{Yin-Zhe Ma$^{1,2,\dagger}$, Enzo Branchini$^{3,4,5,\ddagger}$ \& Douglas Scott$^{1,\star}$\\
$^1$Department of Physics and Astronomy, University of British Columbia, Vancouver, V6T 1Z1, BC Canada.\\
$^2$Canadian Institute for Theoretical Astrophysics, Toronto, Canada.\\
$^3$INAF - Osservatorio Astronomico di Brera, Via Bianchi 46, I-23807 Merate (LC), Italy.\\
$^4$Dipartimento di Fisica, Universit`a degli Studi ¡°Roma Tre¡±,
via della Vasca Navale 84, I-00146 Roma, Italy.\\
$^5$INFN, Sezione di Roma Tre, via della Vasca Navale 84, I-00146,
Roma, Italy.\\
emails: $^{\dagger}$mayinzhe@phas.ubc.ca;\,
$^{\ddagger}$branchin@fis.uniroma3.it;\,
$^{\star}$dscott@phas.ubc.ca}

\maketitle

\begin{abstract}

We constructed a Bayesian hyper-parameter statistical method to
quantify the difference between  predicted velocities derived from
the observed galaxy distribution in the \textit{IRAS}-PSC$z$
redshift survey and  peculiar velocities measured using different
distance indicators. In our analysis we find that the model--data
comparison becomes unreliable beyond $70 \hmpc$ because of the
inadequate sampling by \textit{IRAS} survey of prominent, distant
superclusters, like  the Shapley Concentration. On the other hand,
the analysis of the velocity residuals show that the PSC$z$
gravity field provides an adequate model to the local,
 $\le 70 \hmpc$, peculiar velocity field.
The hyper-parameter combination of ENEAR,  SN, A1SN and SFI++
catalogues in the Bayesian framework constrains the amplitude of
the linear flow to be $\beta=0.53 \pm 0.014$. For an rms density
fluctuations in the PSC$z$ galaxy
 number density $\sigma_8^{\rm gal}=0.42\pm0.03$, we obtain an estimate of the
 growth rate of  density fluctuations
$f\sigma_{8}(z\sim0) = 0.42 \pm 0.033$, which is in excellent agreement with
independent estimates based on different techniques.

\vskip 0.1 truein


\vskip 0.3 truein
\end{abstract}


\section{Introduction}
\label{vel_intro}

The study of peculiar velocity is a powerful tool to explore the
large-scale structure of the Universe. In the standard
$\Lambda$CDM cosmology, gravitational instability causes the
growth of density perturbations and the emergence of the peculiar
velocity field. In the regime where the density perturbation is
linear, the galaxy peculiar velocity ($\vec{v}_{\rm{g}}$), sourced
by the underlying density field, can be expressed as
\citep{Peebles93}
\begin{equation}
\vec{v}_{\rm{g}}(\vec{x})=\frac{H_{0}f_{0}}{4 \pi }\int
d^{3}\vec{x}^{\prime
}\delta _{\rm{m}}(\vec{x}^{\prime },t_0)\frac{(\vec{x}^{\prime }-\vec{x})}{%
\left\vert \vec{x}^{\prime }-\vec{x}\right\vert ^{3}},
\label{eq:vg0}
\end{equation}%
where $H_{0}=H(t_0)$ is the Hubble parameter at the present epoch,
$f_{0}$ is the present day growth rate (henceforth we drop the
subscript $0$) and $\delta_{\rm{m}}$ is the perturbation to the
underlying dark matter distribution, i.e.
$\delta_{\rm{m}}=(\rho-\overline{\rho})/\overline{\rho}$. Assuming
that  the observable galaxy distribution links to the underlying
dark matter distribution through a linear, deterministic bias
factor, $\delta_{\rm{g}}=b \delta_{\rm{m}}$, one can express the
above equation by substituting the growth rate of density
fluctuations $f$ with the dimensionless parameter $\beta \equiv
f/b$. According to Eq.~(\ref{eq:vg0}) the amplitude of the
velocity field scales linearly with $\beta$ and this parameter
fully characterises the model velocity field \citep{Peebles93}.
The value of $\beta$ can be estimated by comparing the model
velocities predicted from
 Eq.~(\ref{eq:vg0})  to measured peculiar velocities estimated from distance indicators.
A good match between these two vector fields would then constitute
an observational test for the gravitational instability paradigm
and for the $\Lambda$CDM model
\citep{Scoccimarro01,Feldman01,Verde02}. Performing such a test is
the aim of this paper.

Comparisons of $v$--$v$ require all-sky redshift surveys to sample
the mass distribution in the local Universe in a dense and
homogeneous way. For this reason, a large number of studies have
used  the \textit{IRAS} 1.2 Jy and PSC$z$ (Point Source Catalogue)
redshift catalogues \citep{Fisher1995,Saunders00}. The latter,
which we shall use in this work, covers about $85$ per cent of the
sky, contains 12,275 galaxies at a mean distance of $7500 \kms$,
and still represents the densest redshift catalogue available to
date. Indeed, the recent 2MRS (Two Micron All Sky Redshift Survey)
$K_{\rm{s}}=11.75$ catalogue \citep{Huchra12} has a larger depth
and better completeness than PSC$z$, but its sampling within $70
\hmpc$ is sparser.

\textit{IRAS} redshift catalogues have been extensively used to
perform $v$--$v$ comparisons to estimate $\beta$. \cite{Davis96}
compared the gravity field obtained from the 1.2 Jy \textit{IRAS}
redshift survey with the Tully-Fisher (TF)  peculiar velocities of
$\sim 2900$ spiral galaxies in the composite Mark III catalogue.
The presence of systematic discrepancies between the two fields
precluded the possibility of measuring $\beta$. The likelihood
technique developed by  \cite{Willick97,Willick98} was designed to
calibrate separately the different subsamples that constitute the
Mark III catalogue. This allowed these authors to eliminate these
discrepancies and to estimate $\beta \simeq 0.49$ from 838
galaxies with TF distances. An independent comparison performed by
\cite{Costa98} using peculiar velocities measured from the
$I$-band TF survey of field spirals (SFI,
\citealt{Giovanelli97,Haynes99})
 found $\beta\simeq 0.6$.
The same likelihood approach as  \cite{Willick97} was used by
\cite{Branchini01} to compare  $989$ SFI galaxies having
TF-peculiar velocities with the PSC$z$ model velocity field. They
found that the a linear field with $\beta \simeq 0.42$ provides a
good match to observations. \cite{Zaroubi02} applied a different
technique to estimate a continuous velocity field from the
measured velocities of both early and late type galaxies in the
ENEAR \citep{Costa00,Bernardi02,Wenger03}
 and SFI catalogues.
They showed that the  PSC$z$ gravity field matches well this
velocity field for  $\beta \simeq 0.51$.

Not all $v$--$v$ comparisons use TF peculiar velocities.
\cite{Nusser01} compared PSC$z$ with the ENEAR early-type galaxies
having $D_{n}$--$\sigma$ velocities and found  $\beta \simeq 0.5$.
\cite{Radburn-Smith04} and \cite{Turnbull12} considered peculiar
velocities of Type Ia supernovae. They considered two different
compilations of objects and found $\beta \simeq 0.5$ and $\beta
\simeq 0.53$, respectively. More recently,  2MASS (Two Micron All
Sky Survey) galaxies \citep{Skrutskie06} were also used to perform
$v$--$v$ comparison. \cite{Pike05} compared the gravity field
predicted from 2MASS photometry and public redshifts to different
peculiar velocity surveys and found $\beta \simeq 0.49$. Finally,
\cite{Davis11} compared the flow traced by the SFI++ sample of TF
velocities \citep{Masters06,Springob07} to that predicted by the
2MRS sample of galaxies brighter than $K_{\rm{s}}=11.25$ and found
a best fit value $\beta \simeq 0.33$.

All these studies assume that $\beta$ is independent of scale.
Since in the standard gravitational instability framework  $f$ is
a  scale-independent quantity, this assumption implies that galaxy
bias is deterministic and linear. In fact, physical processes
related to galaxy formation and evolution result in a
scale-dependent and stochastic bias on small scales, but have
little impact on  $v$--$v$ analyses in which the distribution of
mass tracers is smoothed out on larger scales.

The scatter among  $\beta$ values obtained from different $v$--$v$
comparisons comes from several sources such as: (i) the use of
different techniques  to predict the gravity field and compare it
with measured velocities; (ii) inadequate modelling of the flow in
high density environment; (iii) the use of different mass tracers
to model peculiar velocities; (iv) the use of different velocity
tracers that preferentially
 sample different environments; (v) the possible systematic errors in the calibration of the
distance indicator; (vi) possible systematic errors in the use of
the distance indicators like the Malmquist bias
 \citep{Malmquist20,Lynden-Bell88a,Hudson94,Strauss95}; and
 (vii) systematic biases in the model gravity field, e.g. the
so-called `Kaiser rocket effect'
\citep{Kaiser89,Nusser11,Branchini12}.

These considerations together with the improved quality of the
peculiar velocity data, better understanding of the systematics
and advances in the modelling techniques, lead us to re-examine
the issue. In this paper we make an extensive comparison between
peculiar velocity data from different distance indicators  and the
PSC$z$ density field \citep{Branchini99}. The rationale behind
using different peculiar velocity catalogues is manifold. First,
by performing independent  $v$--$v$ comparisons we can identify
and correct for systematic errors in the measured velocities.
Second, peculiar velocities from different distance indicators
have different random errors and therefore require different
 Malmquist bias corrections. Third, possible additional sources of systematics
can be identified by analyzing the residuals in the  $v$--$v$
comparisons restricted to the individual catalogues. Finally, we
aim at estimating $\beta $ by performing a joint comparison that
involves different velocity catalogues. For this purpose we will
use a Bayesian `hyper-parameter' method \citep{Lahav00,Hobson02}
designed for this purpose.

This paper is organized as follows. In Section \ref{sec-vmodel} we
present the PSC$z$ galaxy catalogue and the model velocity field
observations. In Section~\ref{velocity-cata1} we describe the four
peculiar velocity catalogues considered in this work and the
strategy we adopt to correct for Malmquist bias. The Bayesian
approach adopted to perform the $v$--$v$ comparison is introduced
in Section~\ref{hyper-parameter1} and the results, including the
estimate of $\beta$ are presented in Section~\ref{result}.
Finally, in Section~\ref{sec-discuss}, we discuss our main
conclusions.

\section{Model peculiar velocities}
\label{sec-vmodel}

\begin{figure*}
\centerline{\includegraphics[bb=0 0 635
338,width=3.2in]{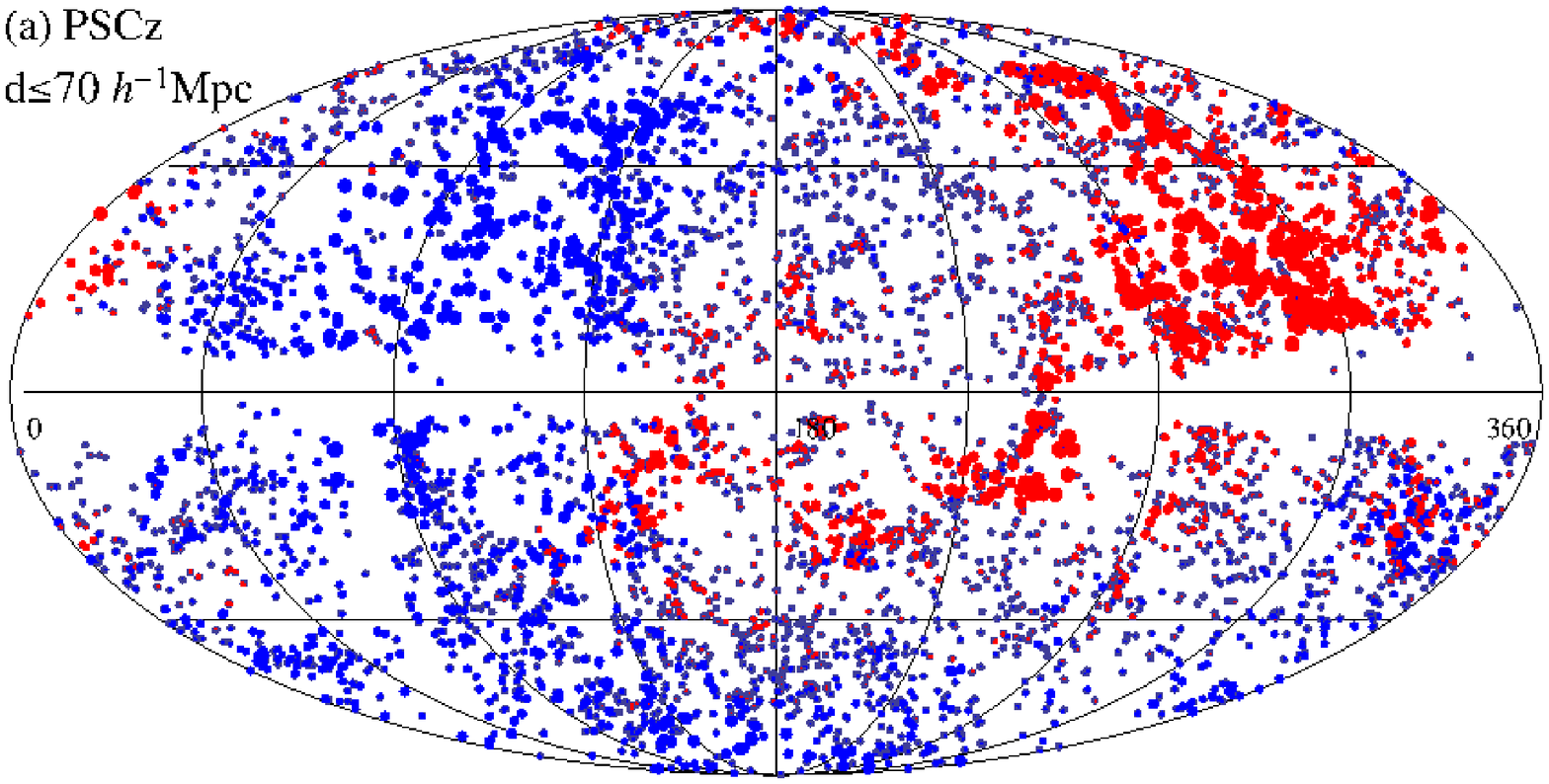}
\includegraphics[bb=0 0 694
367,width=3.2in]{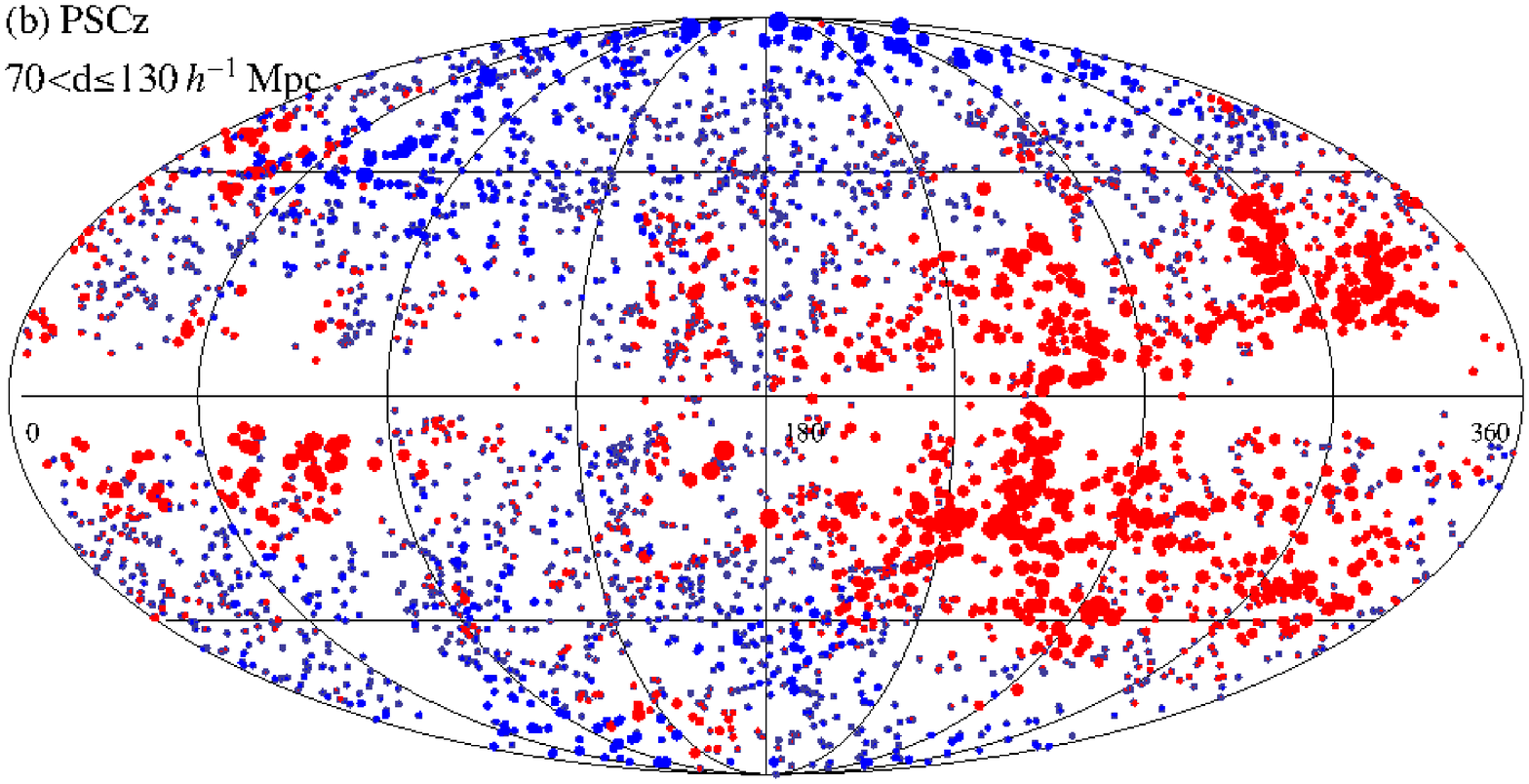}} \centerline{\includegraphics[bb=0
-30 530 270,width=3.2in]{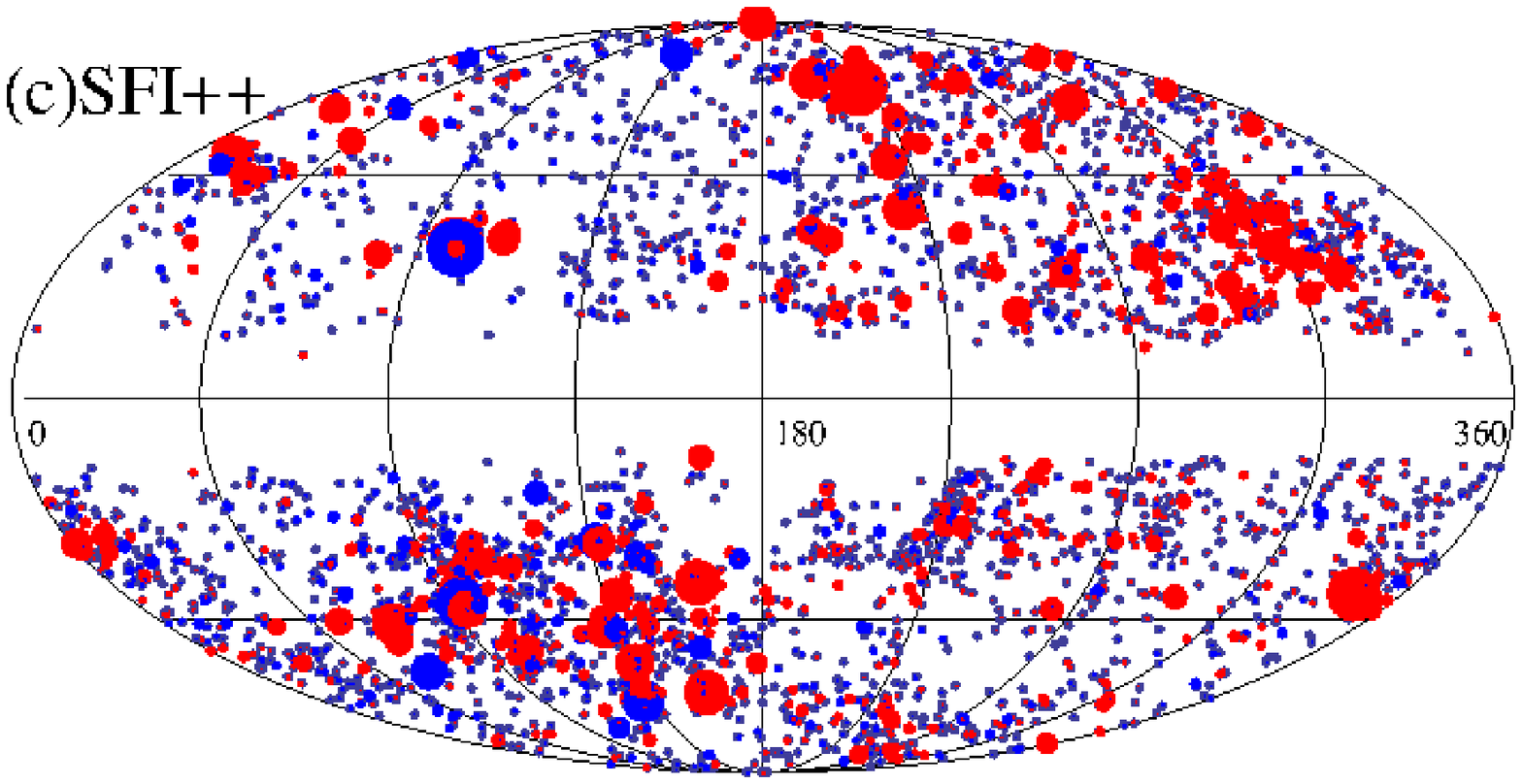}
\includegraphics[bb=0 0
620 403,width=3.0in]{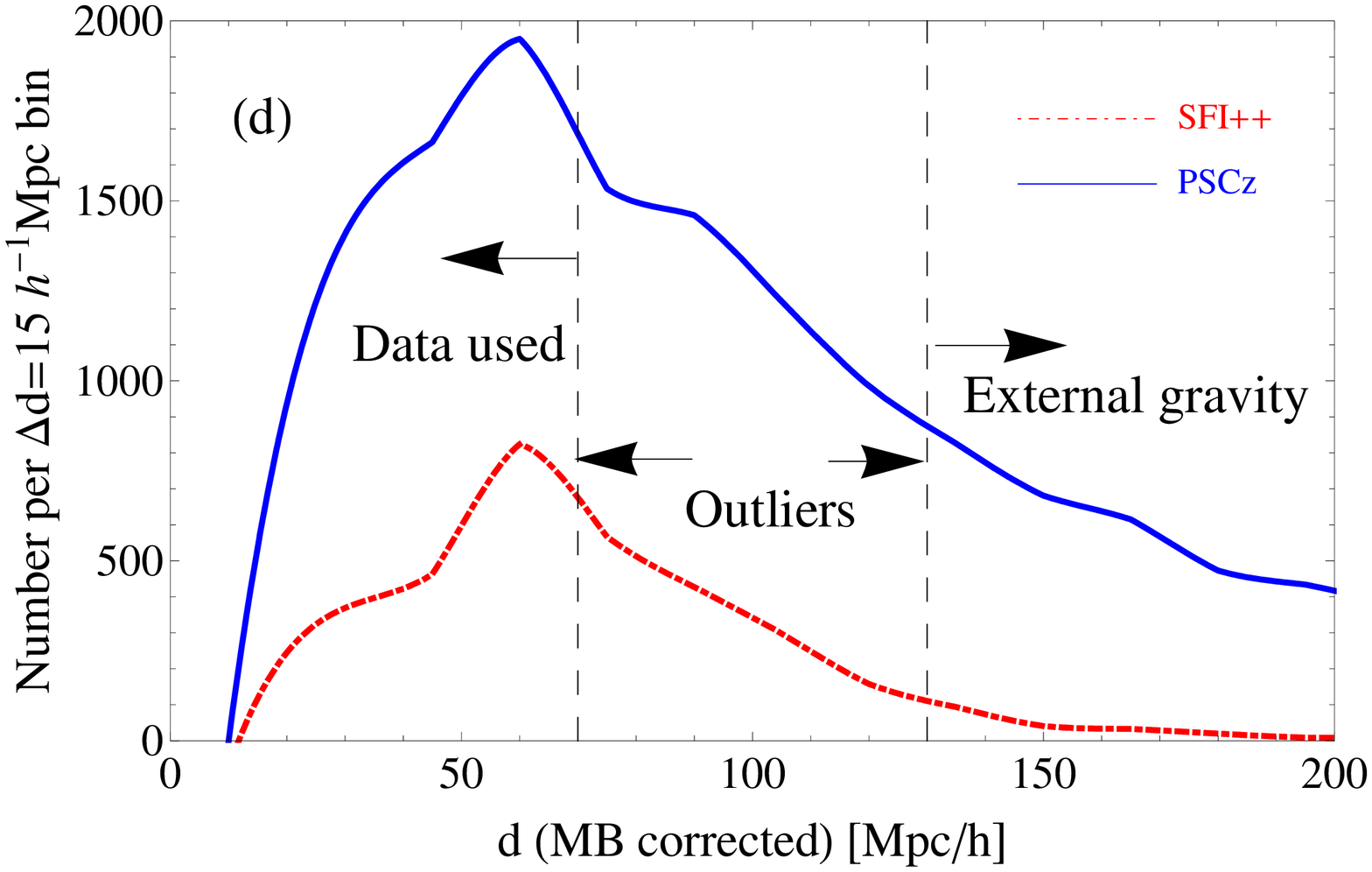}} \caption{(a): full-sky PSC$z$
($d \leq 70 \hmpc$) plotted in Galactic coordinates (5263 samples
in total). The red (light grey) points are moving away from us and
the blue (dark grey) ones are moving towards us. The size of the
points is proportional to the magnitude of the line-of-sight
peculiar velocity. (b): same as (a) but for $70< d \leq 130 \hmpc$
(3732 samples in total). (c): same as (a and b) but for the SFI++
catalogue. (d): number distribution as a function of distance for
PSC$z$ and SFI++ catalogues smoothed over $\Delta d=15 \hmpc$.}
\label{dataplot1}
\end{figure*}

\label{PSCz-describe1}

The first ingredient of  the $v$--$v$ comparison is the model
velocity field. In this work we adopt the one obtained
\citep{Branchini99} from the \textit{IRAS} PSC$z$ catalogue
\citep{Saunders00}. Panels (a) and (b) of Fig.~\ref{dataplot1}
show the angular distribution of PSC$z$ galaxies in Galactic
coordinates. Panel (a) shows objects within a predicted distance
of $70 \hmpc$. This is the distance within which we shall perform
$v$--$v$ comparisons. Panel (b) shows objects in the shell
covering $70$--$130 \hmpc$.

Objects in the PSC$z$ redshift catalogues were used to trace the
underlying mass density field within $300 \hmpc$ under the
assumption of linear and deterministic bias
\citep{Radburn-Smith04}. The model velocity field was obtained
from the positions of galaxies in redshift space according to
Eq.~(\ref{eq:vg0}), using the iterative technique of
\cite{Yahil91}. Iterations involve only objects within $\sim 130
\hmpc$. At larger distances the sampling becomes very sparse and
objects are used to model an external gravity field, but their
peculiar velocities are not used directly.

The iterative procedure is potentially prone to systematic errors.
Here is a short description of the main ones and how to fix them:
\begin{enumerate}

\item Incomplete sky coverage. The surface density of observed
galaxies drops abruptly near the Galactic Plane, in the so-called
Zone of Avoidance (hereafter ZoA), clearly seen  in panels (a) and
(b) of Fig.~\ref{dataplot1}. Different techniques have been
proposed to fill this region and their impact on the final
predictions is not large. \cite{Hudson94} and
\cite{Radburn-Smith04} estimated that typical
 induced errors  on $\beta$ are well below $\sim 8$ per cent, an upper limit obtained by filling
 the ZoA with a homogeneous distribution of fictitious galaxies.
Here we use the filling method of \cite{Branchini99} that consists
of randomly cloning the observed galaxy distribution from the
nearby areas into the ZoA. The results obtained with this method
were almost identical to those obtained using a Fourier-based
technique \citep{Branchini99} and consistent with those of the
Wiener-like filtering analysis of \cite{Zaroubi02}. Analyses
performed using mock PSC$z$ galaxy catalogues showed that the
filling method used in this work induces a spurious bulk flow of
$\sim 60 \kms$ but has negligible impact on the estimated $\beta$
\citep{Branchini99}.

\item Uncertainties in modelling the mass distribution beyond $130
\hmpc$. These uncertainties mainly induce a dipole-like external
field that we remove by computing peculiar velocities relative to
the central observer (i.e.  in the Local Group frame) and by
restricting quantitative analyses to objects within  $70 \hmpc$.

\item Treatment of high density regions. \textit{IRAS} galaxies
are preferentially late type and therefore undersample the cores
of galaxy clusters. In addition, regions around high density peaks
are `tripled valued', i.e. the same redshift is observed at three
different positions along the line of sight to the peak. To
correct these effects we assign an appropriate statistical weight
to all galaxies near the location of known nearby clusters and use
the `robust procedure'  of  \cite{Yahil91} to collapse them to
cluster's centres.

\item Linear scaling, $ \vec{v}(\beta) \propto \beta$ in
Eq.~(\ref{eq:vg0}). Deviations from the simple linear scaling are
observed if peculiar velocities are predicted directly from galaxy
redshifts \citep{Davis11}. However, in the framework of iterative
techniques, the linear scaling is a good approximations and allows
one to reconstruct peculiar velocities to high accuracy
\citep{Yahil91,Branchini99}.

\item Kaiser `rocket' effect. Since the selection function of the
catalogue is initially computed in redshift space, the
reconstruction procedure is potentially prone to the so-called
`rocket' effect \citep{Kaiser89,Branchini12}. Like in the previous
case, the iterative nature of the reconstruction procedure
alleviates the impact of the effect that was corrected by using
mock galaxy catalogues.

\item Non-linear effects. Since the model peculiar velocities are
reconstructed assuming linear theory, non-linear motions need to
be filtered out. Non-linear velocities arise first on small
scales. Effective removal is obtained by filtering the gravity
field on a scale $R_j$, comparable to the mean galaxy-galaxy
separation, while modelling the velocity field at each step of
iterations. The value of $R_j$ determined by \cite{Branchini99} is
shown in figure 3 of their paper. After iterations, the model
velocity field was further smoothed with a top hat filter of
radius $5 \hmpc$ to obtain a uniformly smoothed model velocity
field.

\item Scale-dependent bias. In this work we assume that $\beta$ is
scale independent, i.e. that \textit{IRAS} galaxies trace the
underlying mass density field according to a simple liner relation
on scales larger than $R_j$. This assumption has been explicitly
checked by \cite{Willick97} and \cite{Branchini01}. Reducing $
R_j$ from 5 to $3 \hmpc$, effectively probing scales in which the
bias is non-linear, caused a modest ($\sim 5$ per cent) increase
of $\beta$.

\end{enumerate}

The final result is a linear model for the peculiar velocity field
specified at the reconstructed real space position of 8,995 PSC$z$
galaxies within $130 \hmpc$ that were not collapsed into galaxy
clusters. Their distribution as a function of distance is plotted
in panel (d) of Fig.~\ref{dataplot1} (blue curve). The vertical
line at $70 \hmpc$ indicates the volume considered in the $v$--$v$
comparisons (`Data used' in Table~\ref{tab1}) and the vertical
line at $130 \hmpc$ separates objects with predicted velocities
not used in the quantitative analysis (`Outliers') from those used
to model the gravitational pull of distant structures.

To compare predicted and observed velocities we need to
interpolate model velocities at the positions of galaxies in the
peculiar velocity catalogues. This is done by applying a Gaussian
kernel of the same radius $R_{j}$ to the predicted 3D velocity
specified at the position of the PSC$z$ galaxies,
$\vec{v}_{\rm{rec}}(\vec{x}_{j})$, i.e.
\begin{equation}
\vec{v}_{\rm{smo}}(\vec{x}_{i})=\frac{\sum_{j=1}^{N^{\prime }}\vec{v}_{\rm{rec}}(\vec{x%
}_{j})\exp \left( -\frac{(\vec{x}_{j}-\vec{x}_{i})^{2}}{2R_{j}^{2}}\right) }{%
\sum_{j=1}^{N^{\prime }}\exp \left( -\frac{(\vec{x}_{j}-\vec{x}_{i})^{2}}{%
2R_{j}^{2}}\right) },
\end{equation}
where the sum runs over  the $N'$ PSC$z$ galaxies and
$\vec{x}_{i}$ is the position of the galaxy in the peculiar
velocity catalogue. Interpolated velocities are projected along
the line of sight,
$v_{i,\rm{smo}}=\vec{v}_{\rm{smo}}(\vec{x}_{i})\cdot \hat{r}_{i}$,
in order to be compared to the measured ones.

After correcting for systematic errors, the typical random errors
on predicted velocities, estimated from mock PSC$z$ galaxy
catalogs, is $\sim 130 \ \kms$  \citep{Branchini99}. These errors
are much smaller than those in the
 measured velocities and therefore will be neglected in the $v$--$v$ comparison
presented in this work.

\section{Observed peculiar velocities}
\label{velocity-cata1} Measured peculiar velocities are the other
ingredient of the $v$--$v$ comparison. In this work we consider
four catalogues (ENEAR, SN, SFI++ and A1SN) that we briefly
describe below. We restrict  our attention to these catalogues for
two reasons. First,  they are high quality, recently assembled
data sets that densely sample the peculiar velocity field in the
local Universe. Second, peculiar velocities are estimated from
different distance indicators. These two features minimise the
chance of systematic errors and thus allow us to perform a joint
analysis, reducing random errors considerably.

There are additional velocity catalogues (SC,
\citealt{Giovanelli98,Dale99}), (SMAC,
\citealt{Hudson99a,Hudson04}), (EFAR \citealt{Colless01}), and
(Willick \citealt{Willick99}) that have been recently used to
estimate the bulk flow in the local Universe (e.g.
\cite{Watkins09, Feldman10}). While these other catalogues can be
useful to measure averaged quantities like the bulk flow, they are
too sparse and noisy  to improve the constrains on $\beta$ from a
point-by-point comparison, like in the $v$--$v$ analysis. To
verify this prejudice we did include them in our analysis and
found no improvement in the $\beta$ estimate. For this reason they
are not included in the analysis presented here. The four
catalogues we use are described as follows:

\begin{enumerate}

\item ENEAR. This is a survey of Fundamental Plane distances to
nearby 697 early-type galaxies
\citep{Costa00,Bernardi02,Wenger03}, either isolated or in groups.
Typical errors for the isolated objects are $\sim 18$ per cent of
their distance. The characteristic depth\footnote{The
characteristic depth $\overline{r}$ of each catalogue is defined
as the error-weighted depth
$\overline{r}=\sum_{n}w_{n}r_{n}/\sum_{n}w_{n}$, where
$w_{n}=1/(\sigma^2_{n})$, $\sigma_{n}$ is the measurement error of
line-of-sight velocity
\citep{Watkins09,Feldman10,Ma11b,Turnbull12}} of this sample is
$29 \hmpc$.

\item SN. This sample consists of 103 Type Ia supernovae taken
from the compilation of \cite{Tonry03}. These objects are good
standard candles and their distances are estimated more accurately
($\sim 8$ per cent) than in the previous case. The characteristic
depth of the survey is $32 \hmpc$.

\item SFI++. This is the largest and densest survey of peculiar
velocities available to date \citep{Springob07}. The sample
considered here consists  of  3456 late-type galaxies with
peculiar velocities derived from the Tully-Fisher relation. The
majority of these objects are in the  field (2675) and the rest
found in groups (726). Their distribution across the sky is
remarkably homogeneous, as shown in panel (c)  of
Fig.~\ref{dataplot1}. Their radial distribution (red curve in
panel (d) of the same figure), looks like a scaled-down version of
that of the PSC$z$ galaxies. The characteristic depth of SFI++ is
around $40 \hmpc$. Typical errors are of the order of $23$ per
cent. Unlike for other cases, the estimated peculiar velocities in
the catalogue have been already corrected for Malmquist bias
\citep{Springob07}.

\item A1SN. This catalogue, also known as the `First Amendment'
supernovae sample, contains 245 Type Ia supernovae
\citep{Turnbull12}. It was obtained by merging three data sets:
(1) a sample of 106 objects  from  \cite{Jha07} and
\cite{Hicken09}, (2) a collection of 113 objects  by
\cite{Hicken09}, and (3) 28 objects  from the `Carnegie Supernovae
project' \citep{Folatelli10}. The characteristic depth of the
whole catalogue is $58 \hmpc$, significantly larger than that of
the other catalogues. Typical errors are of the order of $7$ per
cent.

\end{enumerate}

\begin{table}
\begin{centering}
\begin{tabular}{@{}lll}
\hline
 & $d \leq 70$ & $70 \leq d \leq 130$
\\ \hline ENEAR & $632$ & $65$
\\ SN & $72$ & $27$
\\ SFI++ & $2044$ & $1187$
\\ A1SN & $126$ & $104$
\\\hline
\end{tabular}%
\caption{Peculiar velocity samples. The two columns indicate the
number of samples within the range $d \leq 70 \hmpc$ (`in use')
and $70 < d < 130 \hmpc$ (`outliers').} \label{tab1}
\end{centering}
\end{table}

In Table~\ref{tab1} we list the number of objects is each
catalogue. Only those within  $70 \hmpc$, indicated in column 2,
are considered for the $v$--$v$ comparison. At larger distances
the sampling of the underlying velocity field becomes very sparse
and random errors too large. Indeed, when we include `outliers' in
the range ($70$--$130$) $\hmpc$ (column 3) the scatter among the
$\beta$ values obtained from the different catalogues increases,
hinting at possible systematic errors in either predicted or
measured peculiar velocities. Finally, since we perform  the
$v$--$v$ comparison in the Local Group  frame, velocities in the
catalogues that are provided in the CMB frame are transformed  to
the Local Group frame by subtracting the line of sight component
of the Local Group velocity determined from the CMB dipole $v=611
\kms$ toward $(l,b)=(269^{\circ},+28^{\circ})$ \citep{Scott10}.


%
%
%

\subsection{Malmquist bias correction}

Peculiar velocities  in these catalogues are potentially prone to
Malmquist bias (MB) since they were estimated (1) through
`forward' application of the distance indicator, i.e. by
estimating a distance-dependent quantity  from a
distance-independent one and (2) under the assumption that the
estimated  distance is the the best estimate of the true one
\citep{Strauss95}. Of all catalogues considered here only the
SFI++ corrects for MB \citep{Springob07}. In all other cases we
performed our own correction using the procedure outlined below:

Let us
consider the probability distribution of true distance $r$ given the
distance inferred from the distance indicator $d$ and its error
$\sigma $  \citep{Lynden-Bell88a,Strauss95}:
\begin{eqnarray}
P(r|d)=\frac{r^{2}n(r)\exp\left(-\frac{[\ln(r/d)]^2}{2 \Delta^{2}}
\right)}{\int^{\infty}_{0} dr
r^{2}n(r)\exp\left(-\frac{[\ln(r/d)]^2}{2 \Delta^{2}} \right)},
\label{MBP1}
\end{eqnarray}
where $n(r)$ is the mass density along the radial direction and
$\Delta=(\ln(10)/5)\sigma \simeq 0.46 \sigma$ is the fractional
distance uncertainty of the distance indicators. This conditional
probability function can be used to guess the true distance $r$ of
an object from its estimated distance $d$, if the the density
field along  the line of sight to the objects, $n(r)$, is known
{\it a priori}. A popular approach is to assume that $n(r)$ is
constant and the resulting analytic expression is known as {\it
homogeneous}  MB. However, $n(r)$ is far from being constant, even
on the smoothing scales of the model velocity field, and to
correct for the {\it homogeneous} MB we need to follow a different
strategy.

The key issue is to model $n(r)$. We do this using the very same
velocity model of Section~\ref{sec-vmodel}. More precisely, we use
the real-space reconstructed positions of the PSC$z$ galaxies as
mass tracers to interpolate the mass density field on a cubic grid
of  $192 \hmpc$ and mesh size $1.5 \hmpc$, smoothed with a
Gaussian filter of $5 \hmpc$. The field on the lattice is then
interpolated along the line of sight to each object, galaxies and
Type Ia supernovae alike. The value of  $n(r)$ along the line of
sight is specified at the position of 21 equally-spaced points,
with a binning of $1.5 \hmpc$. Finally, Eq.~(\ref{MBP1}) is used
to predict $r$ from $d$ using a MonteCarlo rejection procedure.

\begin{figure*}
\centerline{\includegraphics[bb=0 0 494
332,width=3.0in]{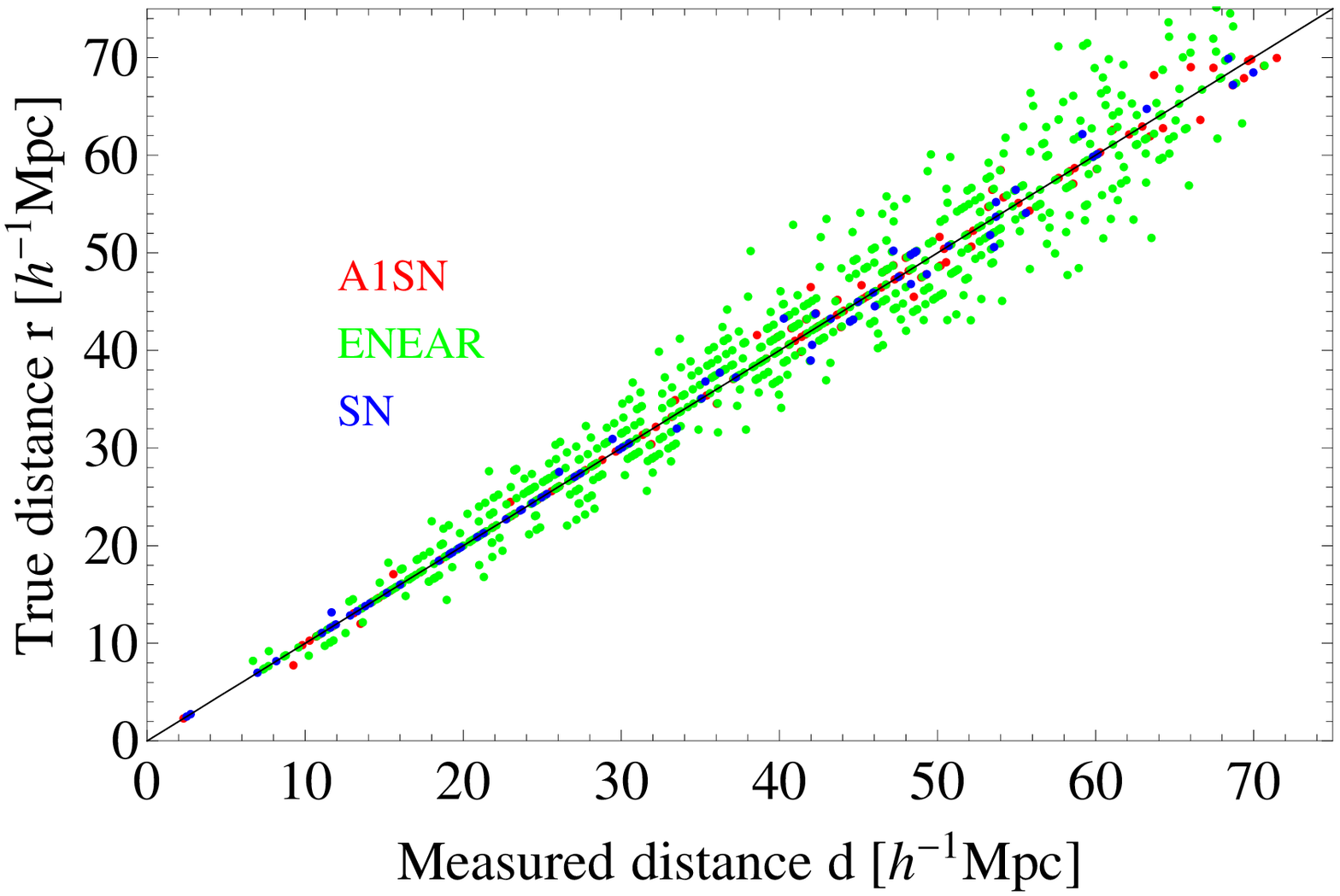}
\includegraphics[bb=0 0 551 350, width=3.2in]{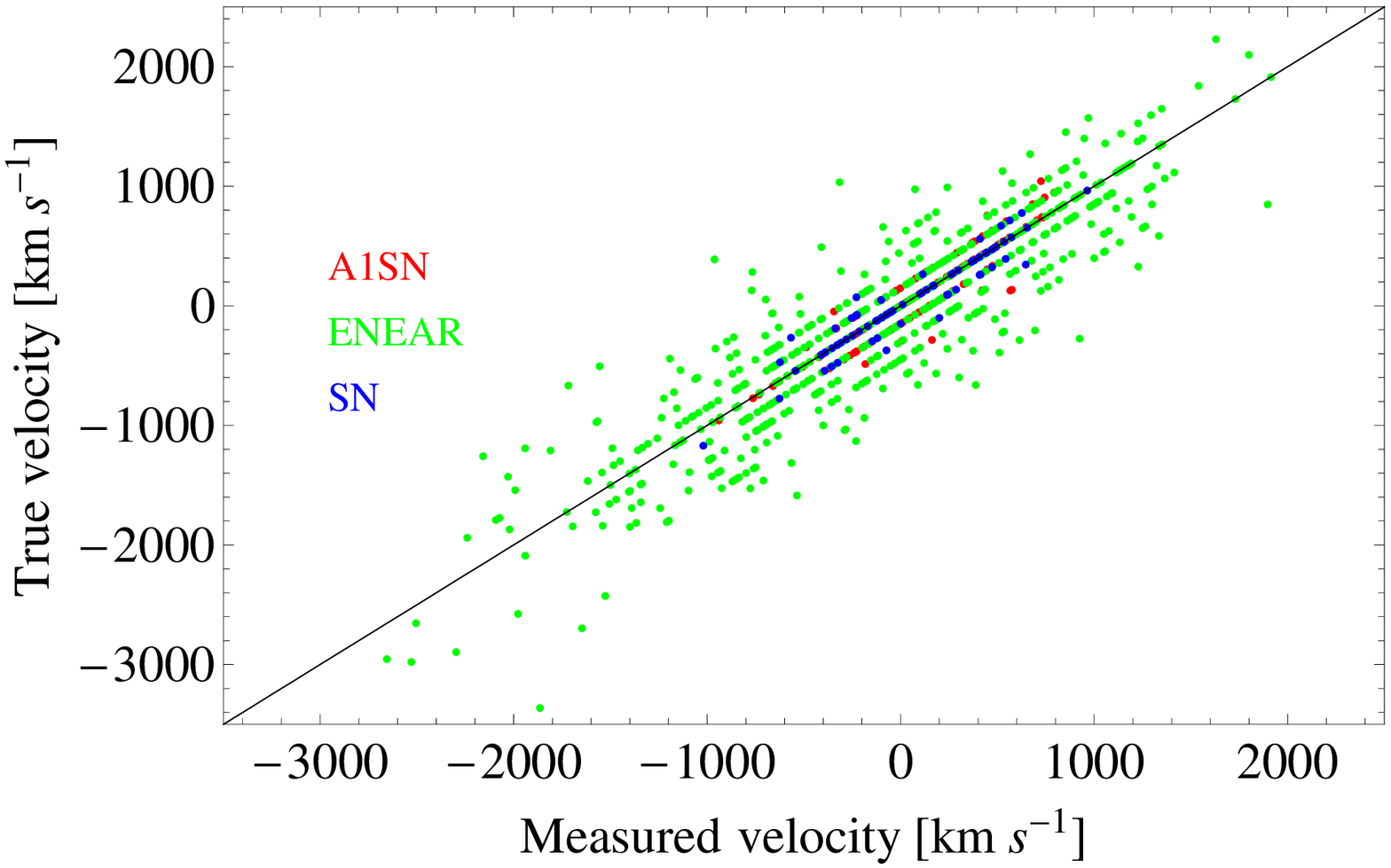}} \caption{Inhomogeneous Malmquist
bias correction. Left panel: the measured distance $d$ (without MB
correction) versus true distance $r$ (MB corrected); Right panel:
measured velocity (without MB correction) versus true velocity (MB
corrected). The $n(r)$ function (Eq.~(\ref{MBP1})) is interpolated
by using PSC$z$ density samples.} \label{MBcorrect}
\end{figure*}

The MB correction is applied to all objects in the catalogues,
apart from SFI++ galaxies. In Fig.~\ref{MBcorrect}, we compare the
measured distance ($x$-axis, before MB correction) and the
estimated true distance ($y$-axis, after MonteCarlo,  MB
correction). Removing this bias preferentially places galaxies at
larger distances. The magnitude of the effect is quantified by the
scatter of the points around the black line. It depends on the
amplitude of the measured velocity error and therefore it
increases with the distance and is smaller for Type Ia SN. The
dispersion around the black line is not symmetric and indicates
that, when averaged over many directions, errors in the observed
velocities preferentially scatter objects to larger distances.

\section{The $v$--$v$ comparison method}
\label{hyper-parameter1}

We are now in the position of comparing observed and model
peculiar velocities at the estimated `true' distances to measure
the value of $\beta$.

We compare the observed peculiar velocities of all objects in all
velocity catalogues with the theoretical predictions. The latter
were basically obtained from Eq.~(\ref{eq:vg0}) and therefore are
sensitive to the mass distribution traced by PSC$z$ galaxies out
to $300 \hmpc$. However, since the selection function of the
catalogue drops beyond $130 \hmpc$, predicted velocities basically
probe the mass distribution (and the value of $\beta$) within this
range. Observed velocities are potentially sensitive to mass
inhomogeneities on much larger scales. However, the fact that we
perform our comparison in the LG frame essentially eliminates the
gravitational pull from scales larger than $130 \hmpc$. Therefore,
the $v$--$v$ comparisons at all points within $70 \hmpc$
effectively probe $\beta$ on scales between the smoothing scale
($\sim 5 \hmpc$ Gaussian) and the effective size of the PSC$z$
survey ($\sim 130 \hmpc$). In this regard, $v$--$v$ comparisons
are complementary to analyses that are based on multipole
decomposition of the observed velocity field (e.g.
\citealt{Watkins09, Feldman10}) which, instead, are sensitive to
the mass distribution on scales larger than that of the peculiar
velocity survey.

The error budget in quantitative $v$--$v$ comparisons is dominated
by uncertainties in the measured velocity. These errors increase
with distance and depend on the distant indicator used and
therefore may vary considerably from catalogue to catalogue. In
addition, different data sets are potentially prone to different
systematic errors in the calibration and in the application of the
distance indicator. If the goal is to perform a joint $v$--$v$
comparison, then these differences need to be properly accounted
for. This can be done by adopting the Bayesian hyper-parameter
method, which is designed to objectively assess whether different
errors in different catalogues are properly accounted for. The
hyper-parameter approach is designed to scale the errors of the
data sets, and then marginalise over all other parameters to
obtain an estimate of the relative statistical `weight' of the
different data sets. In practice, a conventional $\chi^2$ is
defined as $\chi^2=
\sum_{i}(x^{\rm{obs}}_{i}-x^{\rm{the}}_{i}(\theta))^2/(\sigma^2_{i})$,
where $x^{\rm{obs}}_{i}$ and $\sigma_{i}$ are the observed
quantity and its measurement error, and $x^{\rm{the}}_{i}$ is the
corresponding theoretical value with parameters $\theta$. Then the
hyper-parameter effectively scales the errors as $\alpha
\sigma_{i}$. Therefore, by marginalising over the other free
parameters, one can compute the distribution of $\alpha$, which
gives an objective diagnostic of whether the data sets are
problematic and hence deserve further study of the systematic or
random errors \citep{Lahav00}.

Joint analyses of different  velocity catalogues have been
recently performed to investigate the bulk flow, or higher moments
of the cosmic velocity field in the local Universe
\citep{Watkins09,Feldman10}. When combining various data sets the
main issue is the freedom in assigning the relative weights of
different measurements. The standard way of combining two
different data sets (A and B) is by minimising the total $\chi^2$
defined as \citep{Lahav00,Hobson02}
\begin{equation}
\chi^2=\chi^2_{\textrm{A}} + \chi^2_{\textrm{B}}.
\end{equation}
This procedure assumes that one can trust the estimated random
errors, so that the individual $\chi^2$ statistics have equal
weights. However, when combining two or more different data sets
with different errors, one may want to assign different weights to
the individual $\chi^2$ statistics
\begin{equation}
\chi^2=a \chi^2_{\textrm{A}} + b \chi^2_{\textrm{B}},
\end{equation}
where $a$ and $b$ are the Lagrangian multipliers that
constitute the Bayesian hyper-parameters. Therefore,
even if the measurement errors are inaccurate, the
hyper-parameters can assess the relative weight of different
experiments, and hence let the experiments objectively determine
their own weights.


In the Bayesian  hyper-parameter method framework , the a
posteriori distribution of the parameter $\theta$ is defined as
\citep{Lahav00}

\begin{equation}
-2 \ln P(\theta|D) = \sum_{k}N_{k}\ln \chi_{k}^{2},
\label{hyper-definition}
\end{equation}%
where $D$ represents the data, the sum is over all data sets,
$N_{k}$ is the number of data in each data set and $\chi^{2}_k$ is
the $\chi^{2}$ of the $i$th data set (see  \citealt{Hobson02} and
Appendix B of \citealt{Ma10} for detailed discussions).


In this work we define the $\chi^2$ for each velocity catalogue as
\begin{equation}
\chi ^{2}(\beta ,\alpha)_k=\sum_{i=1}^{N_{k}}\left( \frac{%
v^{\rm{mea}}_{i}-\beta \cdot v^{\rm{smo}}_{i}}{\alpha
\sigma^{\rm{mea}}_{i}}\right) ^{2}, \label{chi2}
\end{equation}%
where $\sigma^{\rm{mea}}_{i}$ is the measurement error for the
line-of-sight peculiar velocity $v^{\rm{mea}}_{i}$, and  $\alpha$
is the hyper-parameter of the catalogue. Model velocities
$v^{\rm{smo}}$ are normalised to $\beta=1$ and linearly scaled by
the free parameter $\beta$ according to Eq.~(\ref{eq:vg0}).
Then the likelihood
function becomes
\begin{equation}
\mathcal{L}(\beta,\alpha) \sim
\alpha^{-N_k}e^{-\frac{1}{2}\chi^2_k}.\label{likeli2}
\end{equation}

Note that this is the likelihood of each individual catalogue
characterized by its hyper-parameter $\alpha$ that represents the
(unknown) scaling of measurement error. The distribution of
$\beta$ for the single catalogue can be obtained by marginalise
the  likelihood in Eq.~(\ref{likeli2})  over $\alpha$ and vice
versa.

In contrast, Eq.~(\ref{hyper-definition}) defines the posterior
probability  of the hyper-parameters and $\beta$ given a
combination of different data sets $D$. It is obtained from the
combination of likelihoods of the different catalogues in
Eq.~(\ref{likeli2}). The distribution of $\beta$ from the joint
analysis is obtained by minimising Eq.~(\ref{hyper-definition})
with respect to $\beta$.



\begin{figure*}
\centerline{\includegraphics[bb=0 0 625
418,width=3.2in]{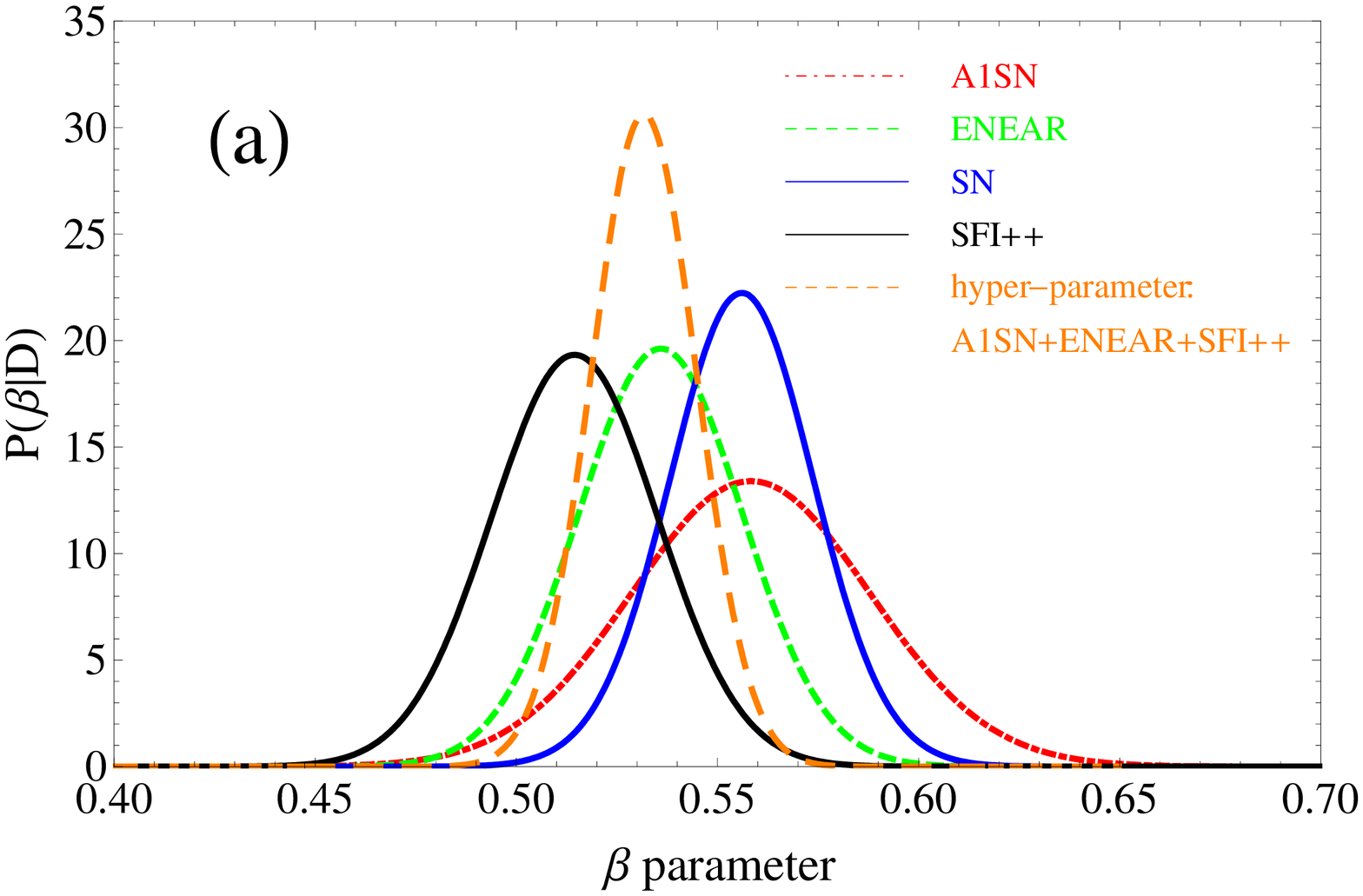}
\includegraphics[bb=0 0 620 398,width=3.2in]{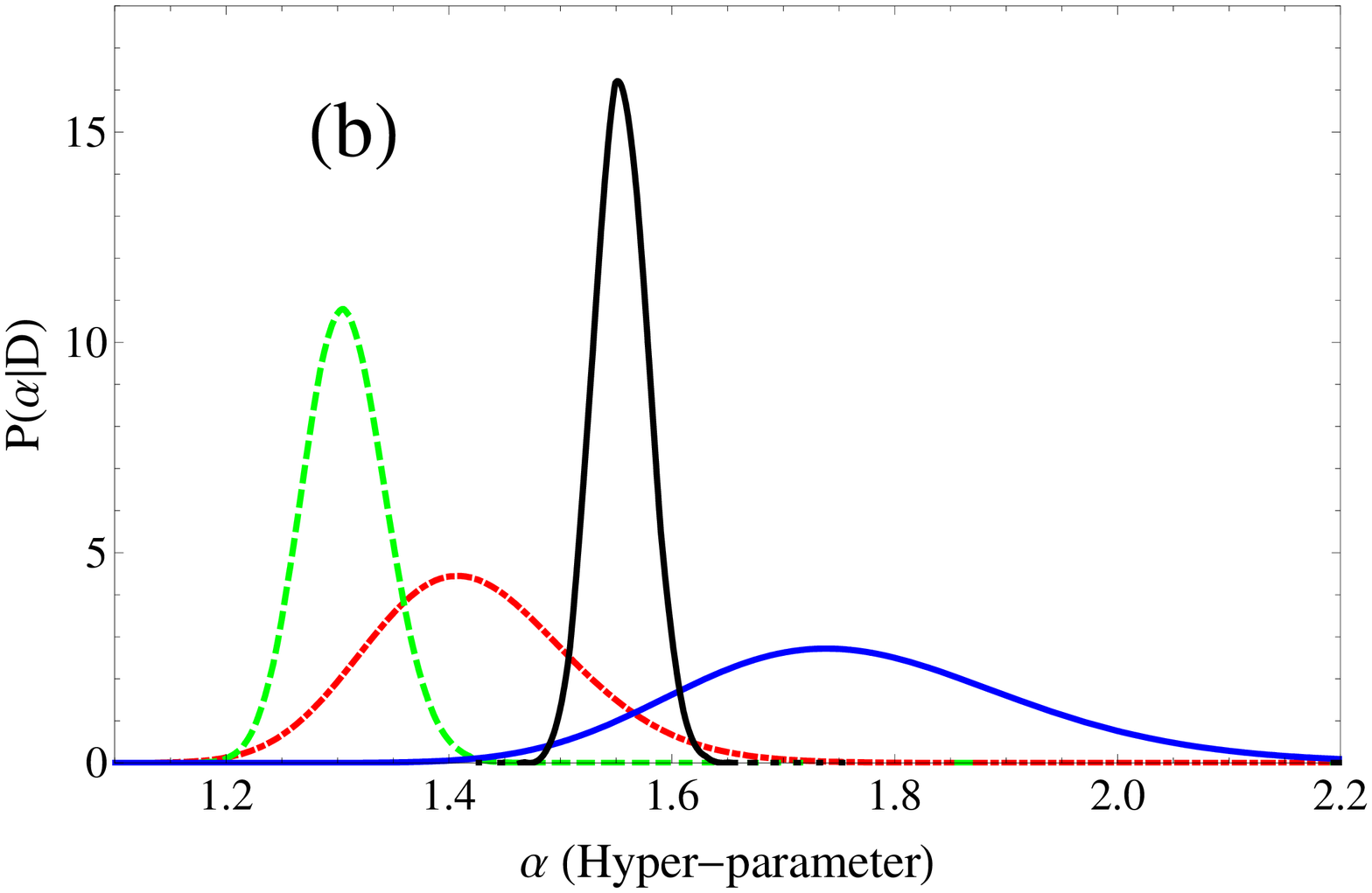}}
\centerline{\includegraphics[bb=0 0 615
398,width=3.4in]{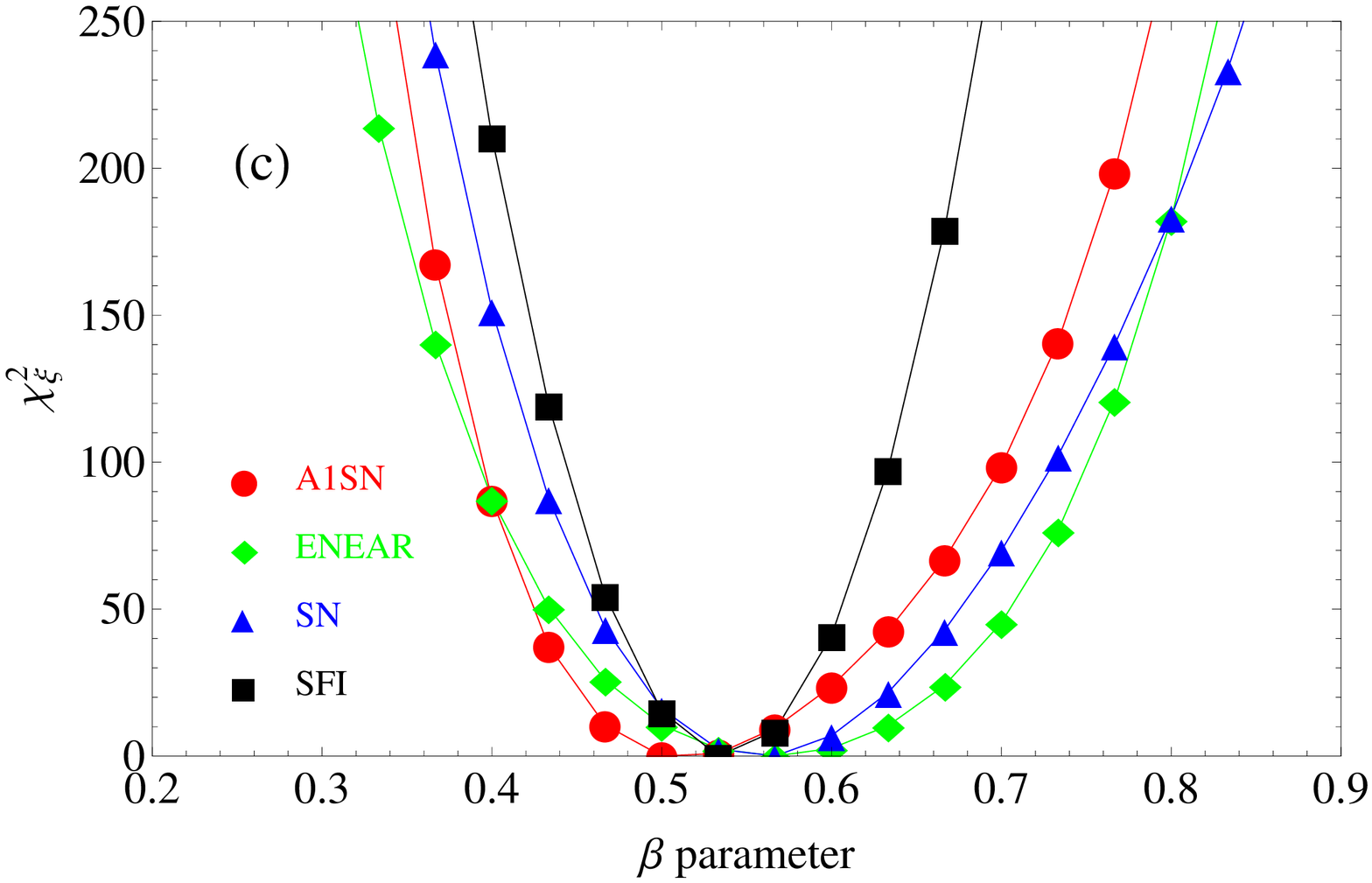}} \caption{(a): Marginalised
posteriori distribution of $\beta$ by applying four different
catalogues ($d \leq 70 \hmpc$). The orange line is the combined
constraint by using the hyper-parameter method. (b): Posteriori
distribution for hyper-parameter $\alpha$, same colour scheme as
panel (a). (c): Goodness of fit for residual velocity correlation
$\chi^2_{\xi}$.} \label{like-plot}
\end{figure*}

\begin{table*}
\begin{centering}
\begin{tabular}{@{}lllll}\hline
 & $\beta$ (likelihood (\ref{chi2}) and (\ref{likeli2})) & $\beta$ (residual velocity)
 &  $\beta$ (reference) & $\alpha$ (this study)
\\ \hline  ENEAR & $0.54 \pm 0.022$ & $0.56 \pm 0.022$ & $0.50 \pm 0.10$ \citep{Nusser01} &
$1.31 \pm 0.04$
\\  SN & $0.56 \pm 0.022$ & $0.56 \pm 0.022$ & $0.55 \pm 0.06$ \citep{Radburn-Smith04} &
$1.74 \pm 0.14$
\\  SFI++ & $0.51 \pm 0.022$ & $0.54 \pm 0.014$ & 
& $1.55^{+0.03}_{-0.02}$
\\  A1SN & $0.56 \pm 0.032$ & $0.51 \pm 0.022$ & $0.53 \pm 0.08$ \citep{Turnbull12} &
$1.41^{+0.09}_{-0.08}$
\\  \textbf{Combined (hyper)}  & $\mathbf{0.53 \pm 0.014}$ & &  &
\\\hline
\end{tabular}%
\caption{Constraints on $\beta$ and $\alpha$ for different
catalogues and combinations. We also list the constraints from
other published studies.} \label{tab2}
\end{centering}
\end{table*}

\section{Results}
\label{result} In this section we present the results obtained
from the hyper-parameter method, investigate  the $v$--$v$
comparisons separately  for each catalogue and assess the goodness
of the fit by analysing the correlation among the residuals in the
comparisons.

\subsection{$\beta$ value and  hyper-parameters}
The results of the hyper-parameter analysis are summarised in the
top panels of Fig.~\ref{like-plot}. Panel (a), on the left,  shows
the posterior probability of the $\beta$ value obtained form each
velocity catalogue  after marginalising over the corresponding
hyper-parameter $\alpha$. Different colours refer to different
catalogues, indicated by the labels. SFI++ galaxies prefer a lower
value of $\beta$, whereas the distribution of both Type Ia
supernovae catalogues peak at larger values. However, the overlap
among the different probability distributions is significant,
indicating  that the  $\beta$ values obtained from the different
$v$--$v$ comparisons agree with each other.

This impression is confirmed in Table~\ref{tab2}, where we list
the values of $\beta$ at the peak of the various distributions,
together with the $\pm 1\sigma$ width of their  Gaussian
distributions (column 2). For comparison we also list the $\beta$
values obtained from $v$--$v$ comparisons based on the same
velocity catalogues. (column 4 in Table~\ref{tab2}). They are
listed only for reference, since a quantitative comparison should
account for the different model velocity fields, comparison
techniques and objects considered in each analysis. No $v$--$v$
comparison was performed with the SFI++ velocities.

In  Fig.~\ref{like-plot} the orange curve that peaks at $\beta
\simeq 0.53$ is the result of combining all catalogues in the
joint hyper-parameter analysis. The fact that the distribution
largely overlaps with those obtained for the individual catalogues
confirms the consistency among the results and allows us to
estimate $\beta$ with a $\sim 2$ per cent error. This constraint
is remarkably tight, but we have to keep in mind that we are
assuming here that errors are solely contributed by uncertainties
in measured velocities, and that all other possible sources,
including cosmic variance, can be safely neglected.

In panel (b) of Fig.~\ref{like-plot}, we plot the marginalised
distribution of hyper-parameter $\alpha$ for the four catalogues.
Their values are listed in column 5 of Table~\ref{tab2}. The fact
that $\alpha=N_k/\chi^2$ is larger than unity for all subsets
suggests  that random errors have been systematically
underestimated by a factor  $\sim \alpha^{1/2}$. In our analysis
we have assumed that the error budget is dominated by
uncertainties in measured peculiar velocities and have ignored
errors in the velocity model. We've justified this assumption in
Section~\ref{sec-vmodel} based on the mock catalogue analysis
performed by
 \citep{Branchini99, Branchini01}.
Another error source that we have neglected is associated to the
procedure adopted to correct for the Malmquist bias. To check
whether this can indeed account for the remaining error we have
performed $1000$ Montecarlo realization of the MB correction and
evaluate its uncertainties from  the scatter in the value of
$\beta$. We have done this exercise for the ENEAR catalogue and
found that the MB correction induces an error
 $\sigma_{\beta} \sim 0.01$ of the same size as the one from measured velocities.
 When the two are added in quadrature the total error increases by a factor
 $\sim 1.4^{1/2}$, consistent with the hyper-parameters valued.

 We conclude that errors on $\beta$ are contributed by uncertainties in the
 measured velocities and in the MB corrections and that, for each velocity catalogue,
 the latter can be estimated scaling the former by the hyper parameter $\alpha$.
 All $\beta$ errors quoted in Table~\ref{tab2} have been estimated in this way.

\subsection{Individual $v$--$v$ comparisons}

\begin{figure*}
\centerline{\includegraphics[bb=0 0 543
359,width=3.0in]{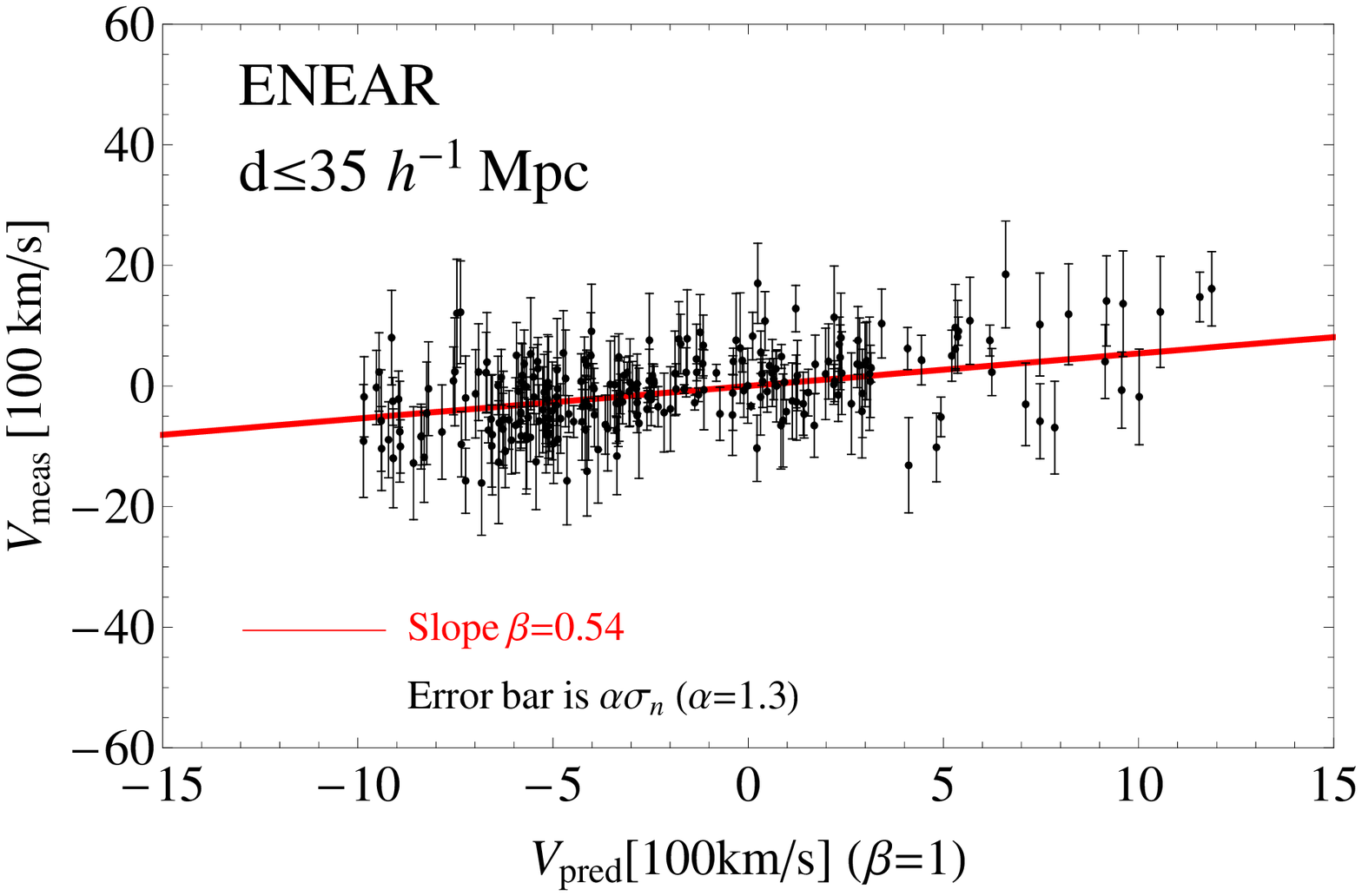}
\includegraphics[bb=0 0 570 353,width=3.1in]{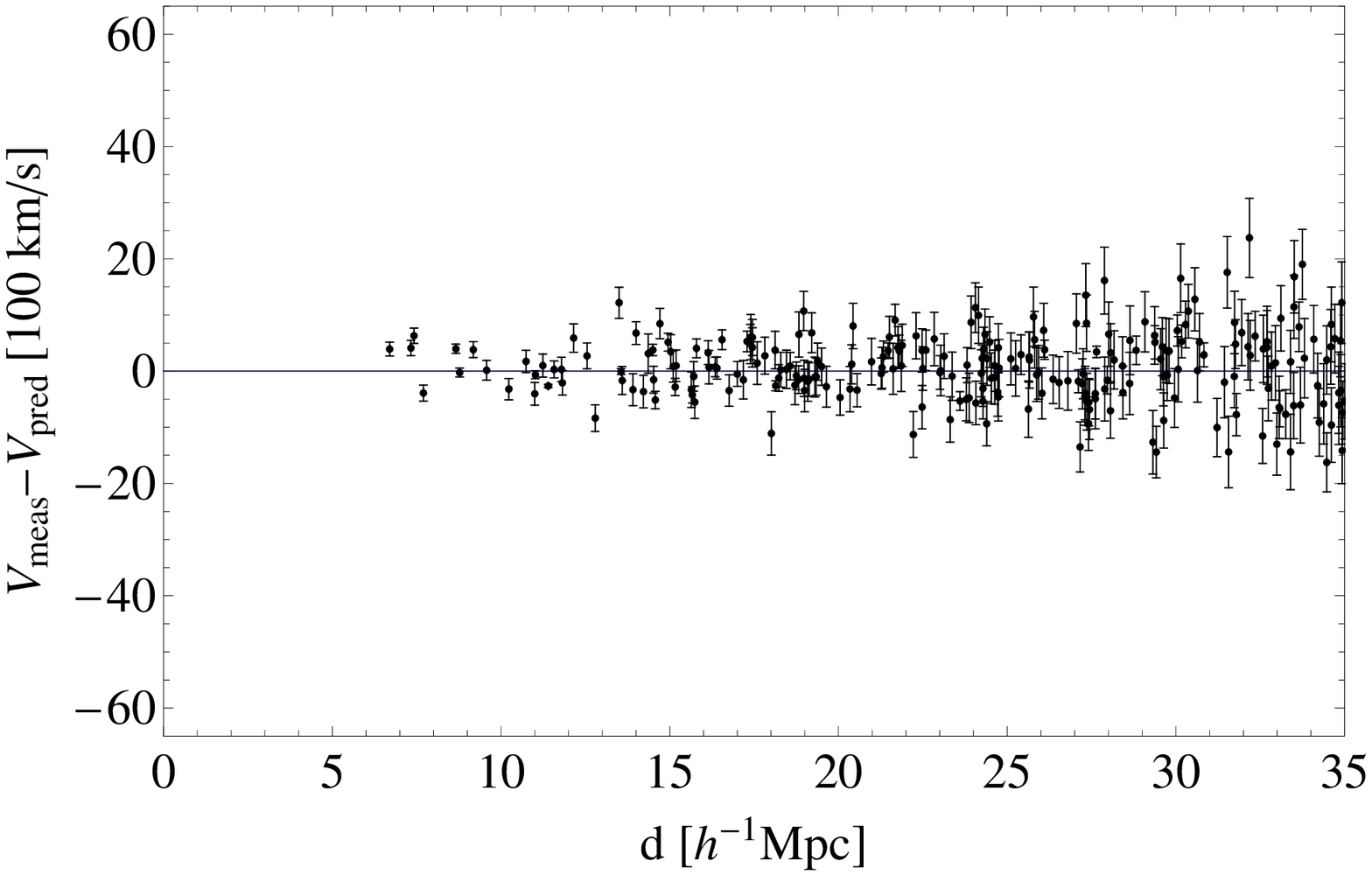}}
\centerline{\includegraphics[bb=0 0 583
388,width=3.0in]{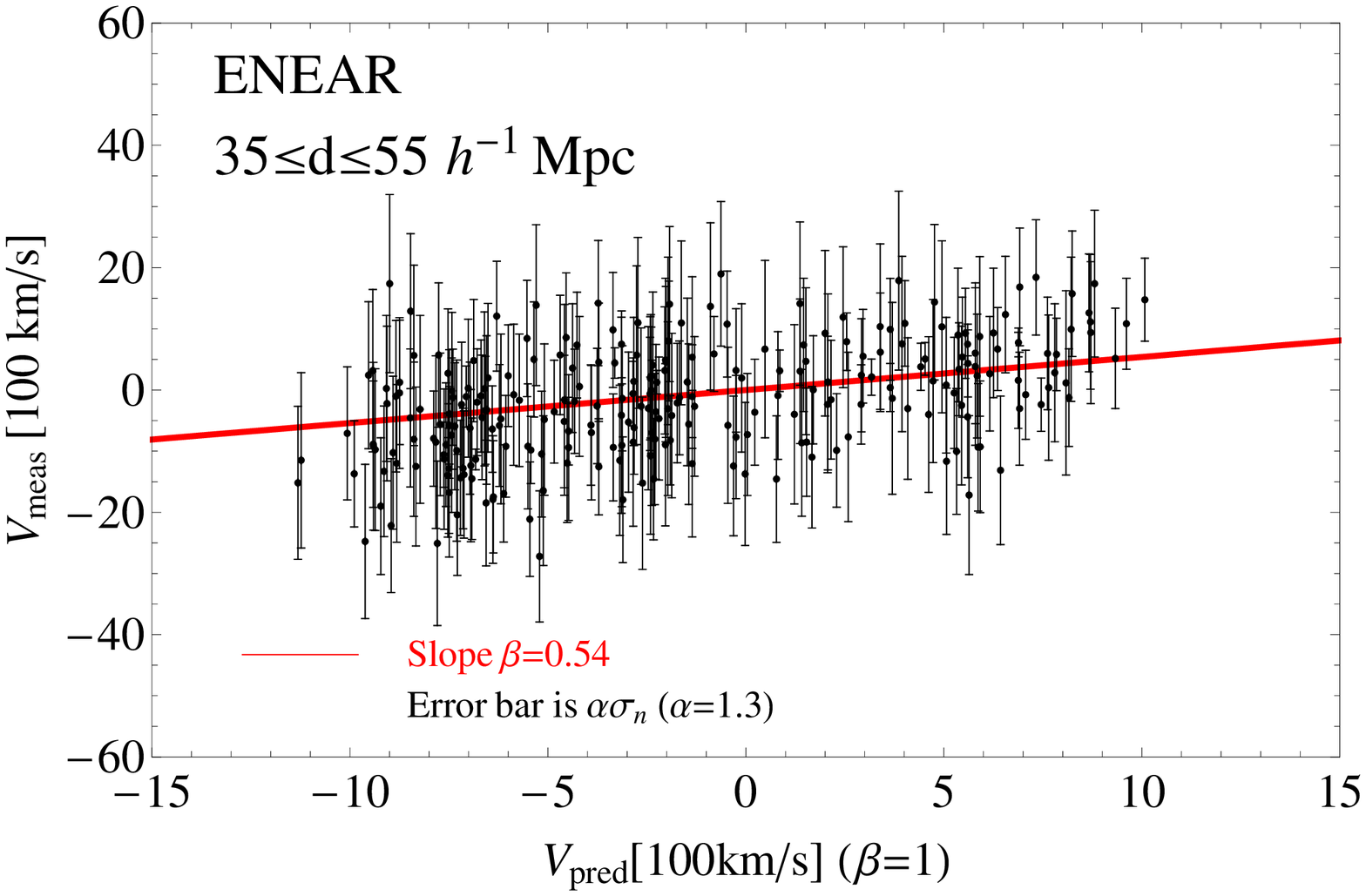}
\includegraphics[bb=0 0 554 364,width=3.0in]{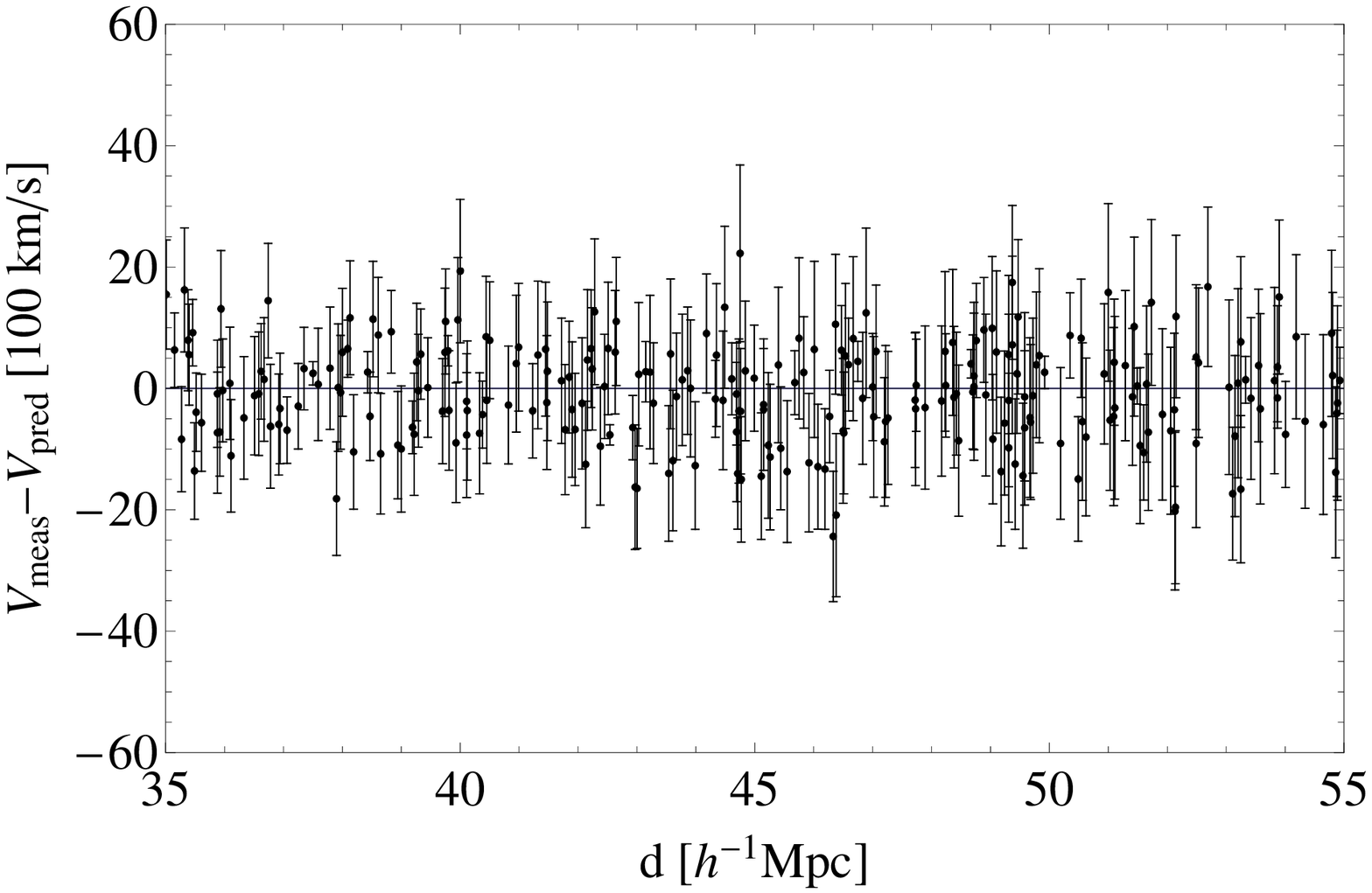}}
\centerline{\includegraphics[bb=0 0 590
386,width=3.0in]{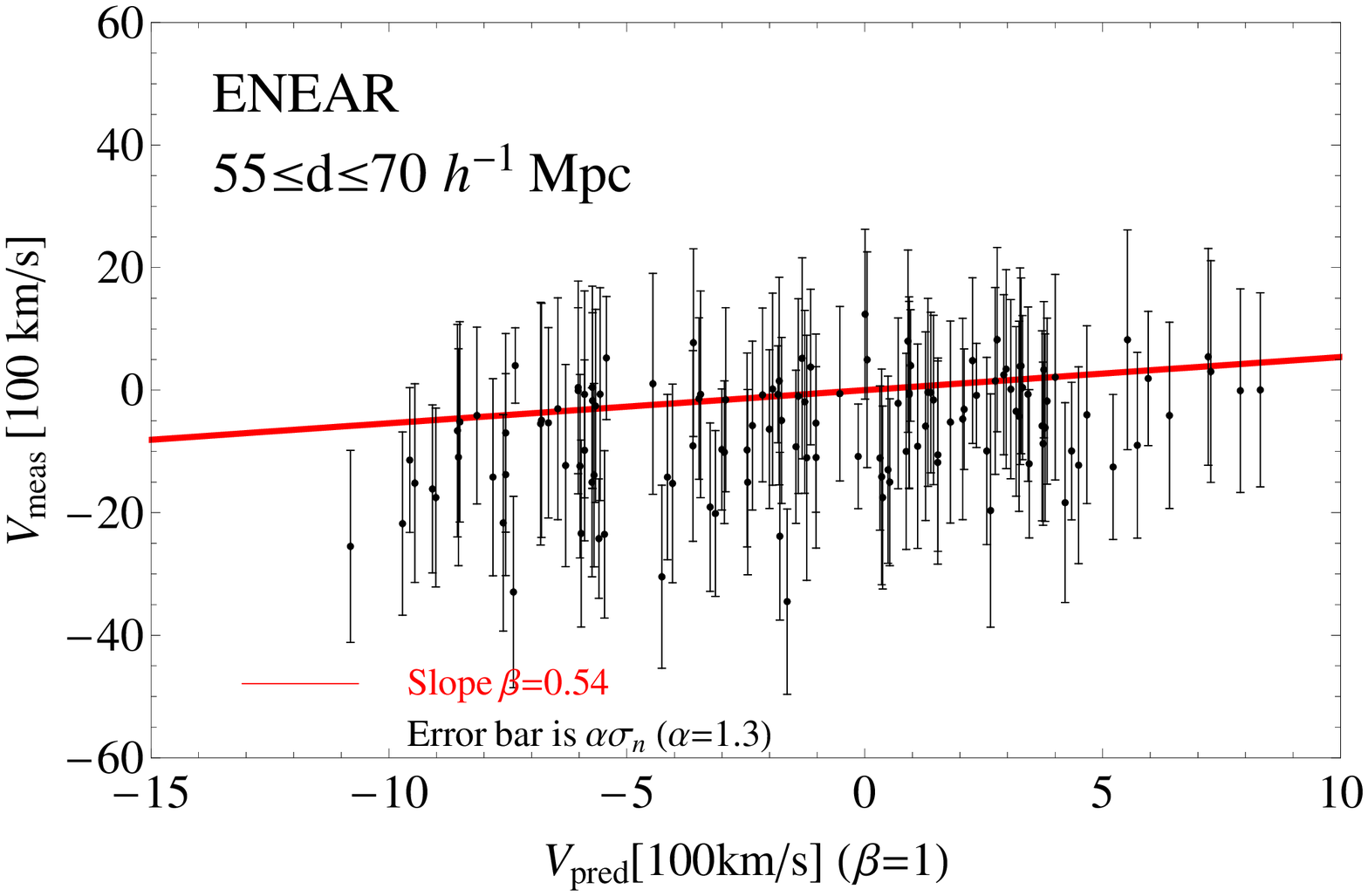}
\includegraphics[bb=0 0 533 347,width=3.0in]{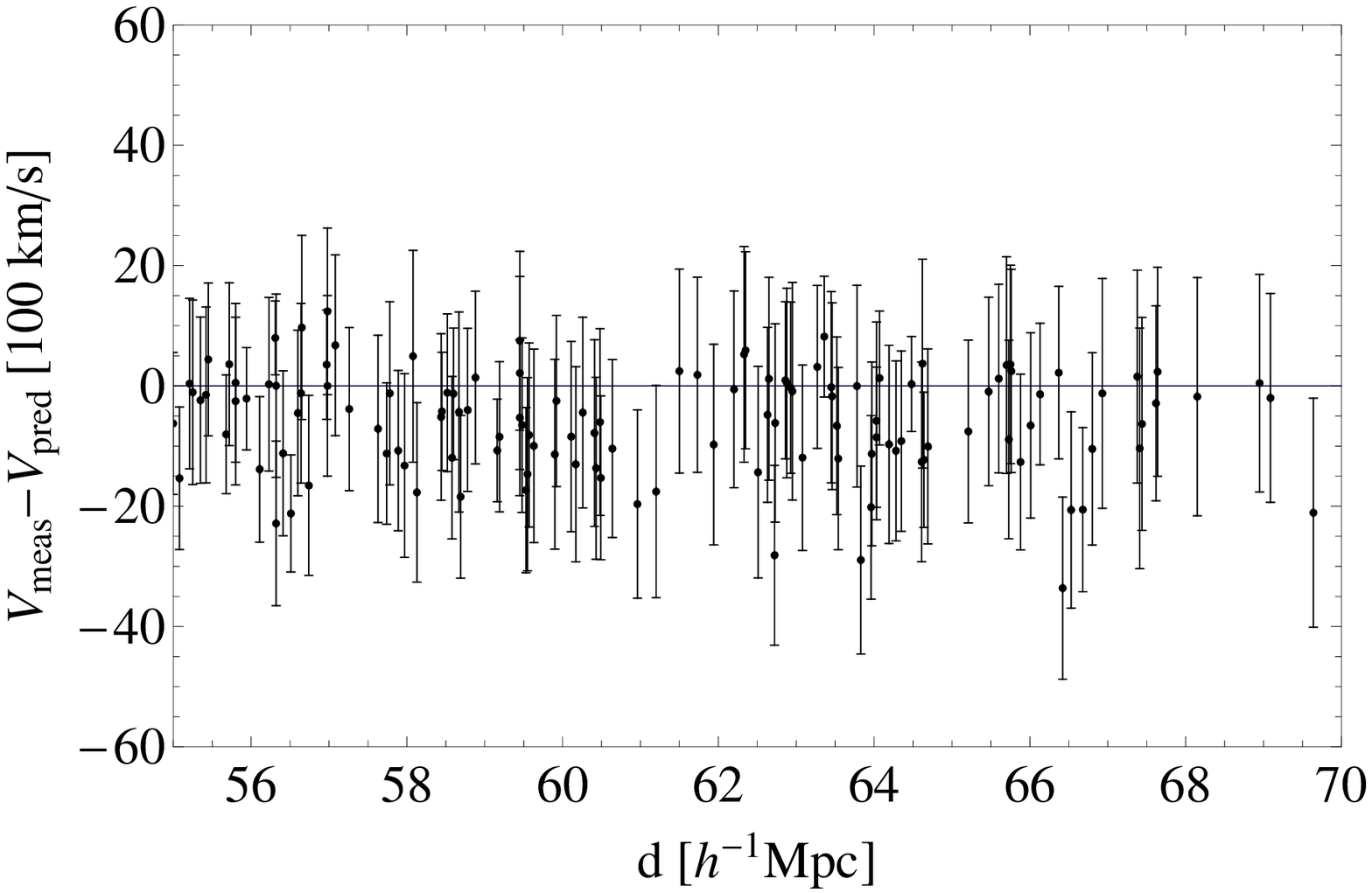}}
\caption{Comparison between measured line-of-sight velocities of
the ENEAR catalogue and the PSC$z$ gravity field. Left column:
direct comparison with $\beta=0.54$ as the best-fit value
(Table~\ref{tab2}). Right column: residual velocities, i.e. the
reconstructed velocities
 subtracted from the measured velocities. All of
the errors here are measurement errors multiplied by the best-fit
value of the hyper-parameter, $\alpha=1.3$ (Table~\ref{tab2}). The
three rows correspond to different distance intervals.}
\label{enearplot1}
\end{figure*}

\begin{figure*}
\centerline{\includegraphics[bb=0 0 485
324,width=3.0in]{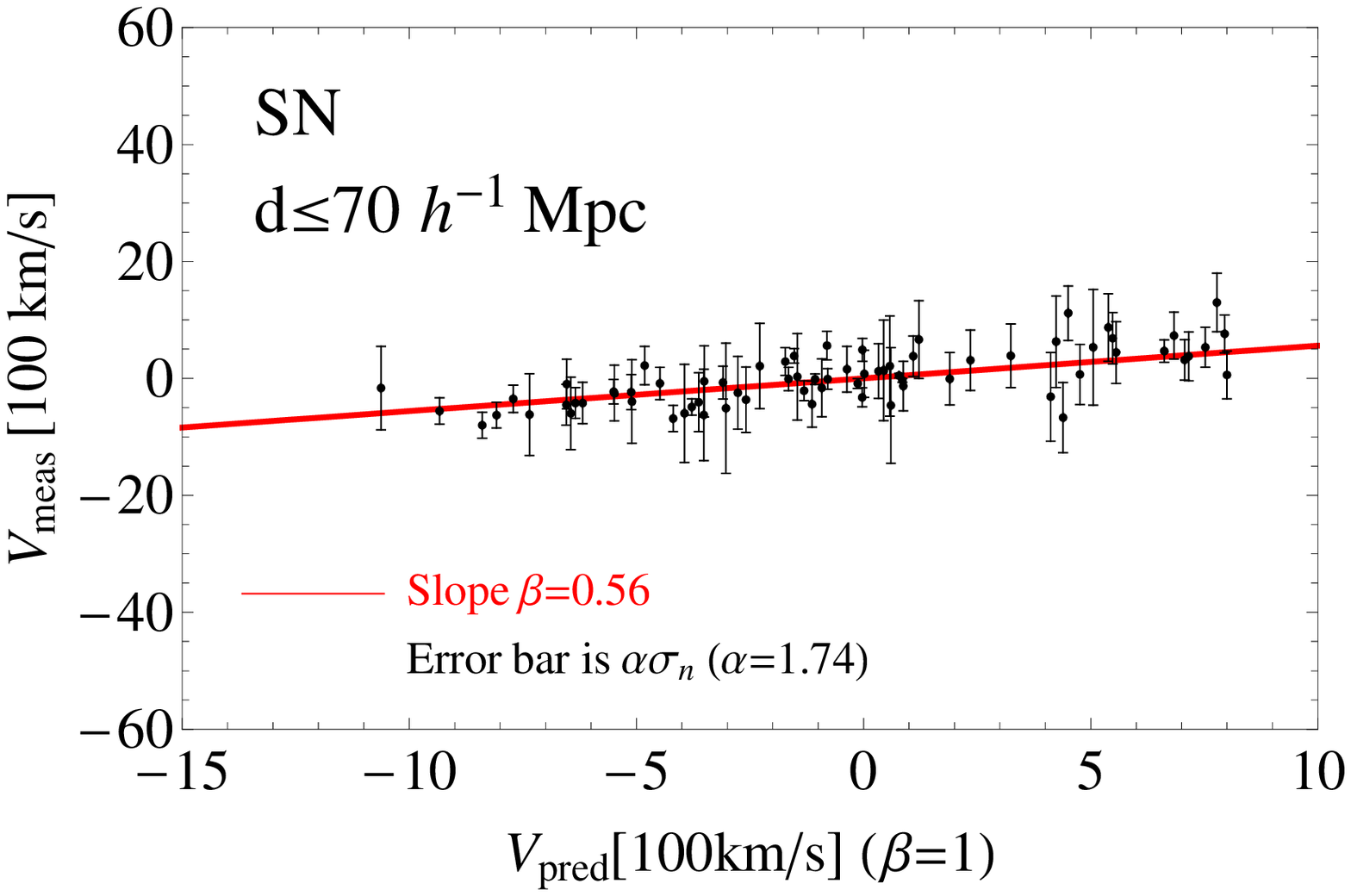}
\includegraphics[bb=0 0 491 320,width=3.0in]{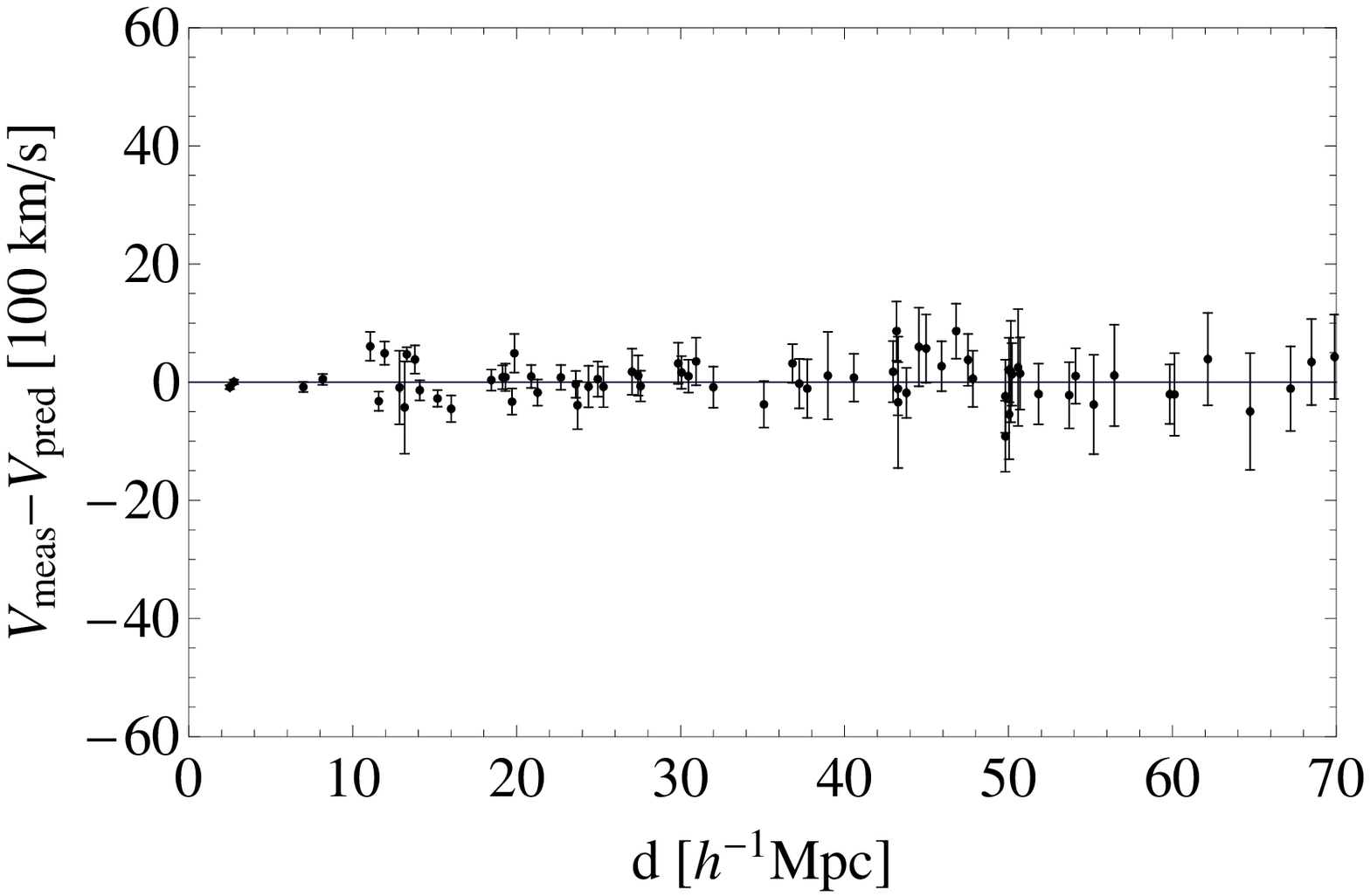}}
\caption{Same as Fig.~\ref{enearplot1} for the SN catalogue.}
\label{snplot1}
\end{figure*}

\begin{figure*}
\centerline{\includegraphics[bb=0 0 515
341,width=3.0in]{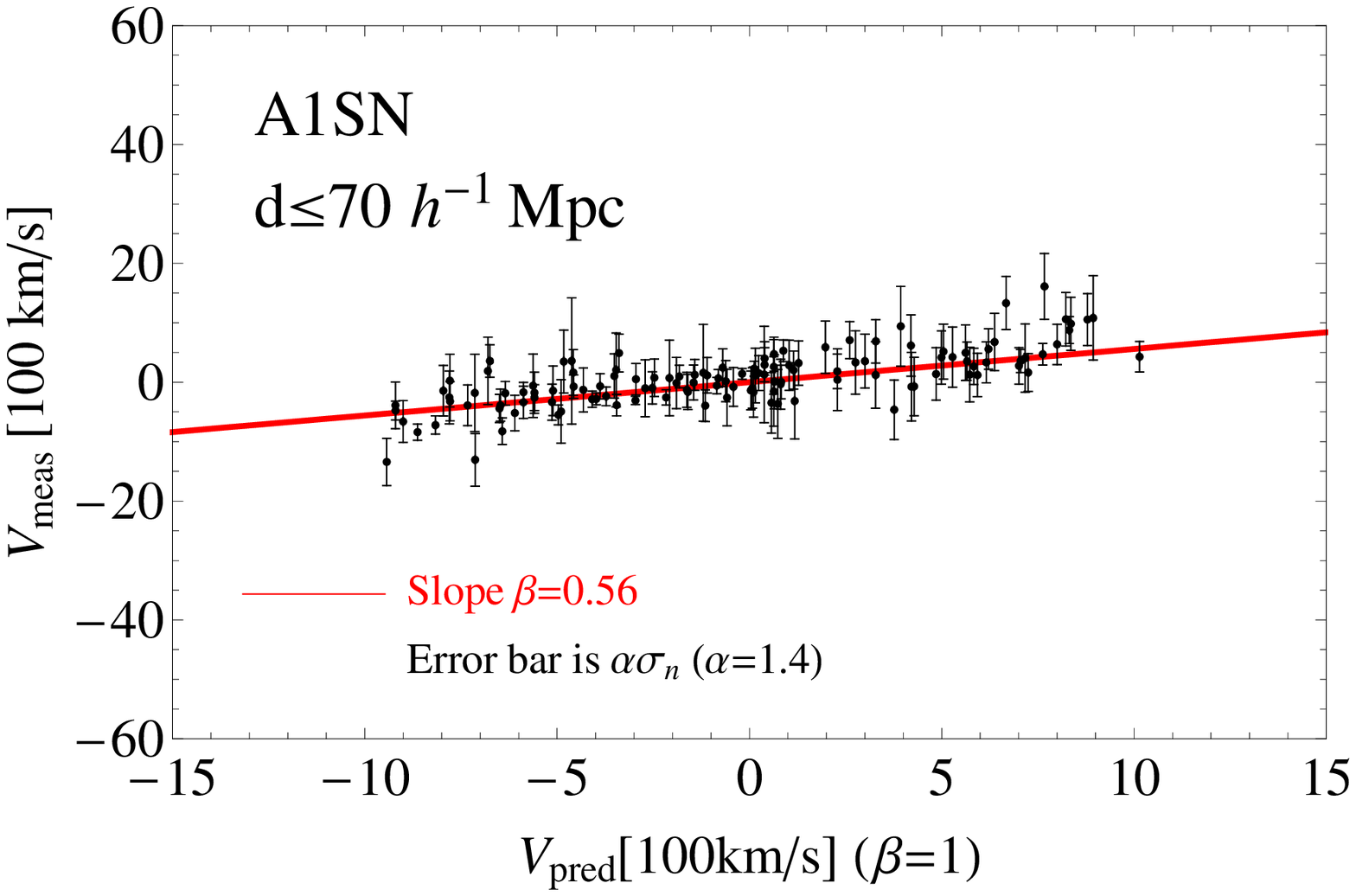}
\includegraphics[bb=0 0 500 329,width=3.0in]{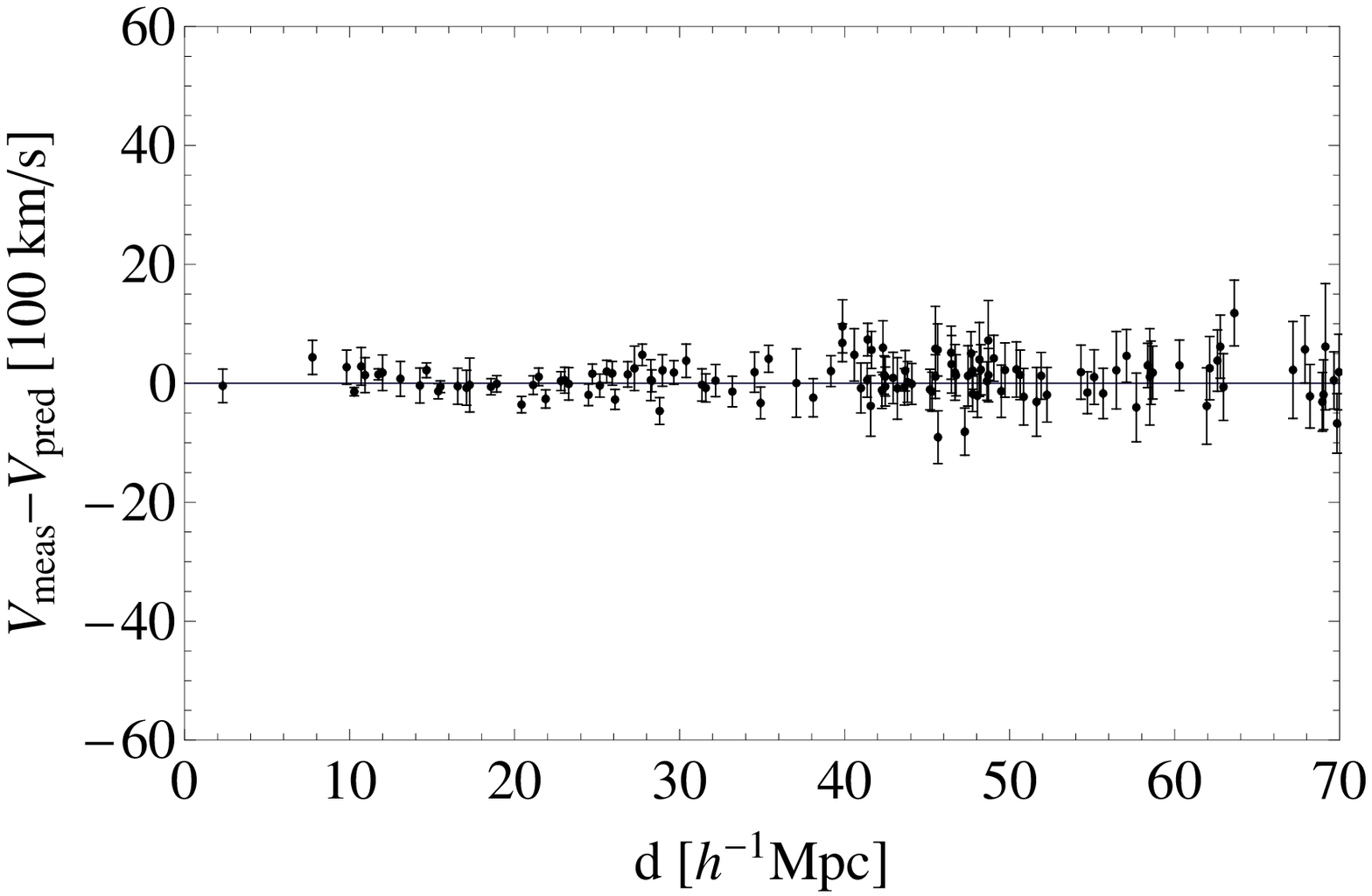}}
\caption{Same as Fig.~\ref{enearplot1} for the A1SN catalogue.}
\label{a1snplot1}
\end{figure*}

\begin{figure*}
\centerline{\includegraphics[bb=0 0 587
389,width=3.0in]{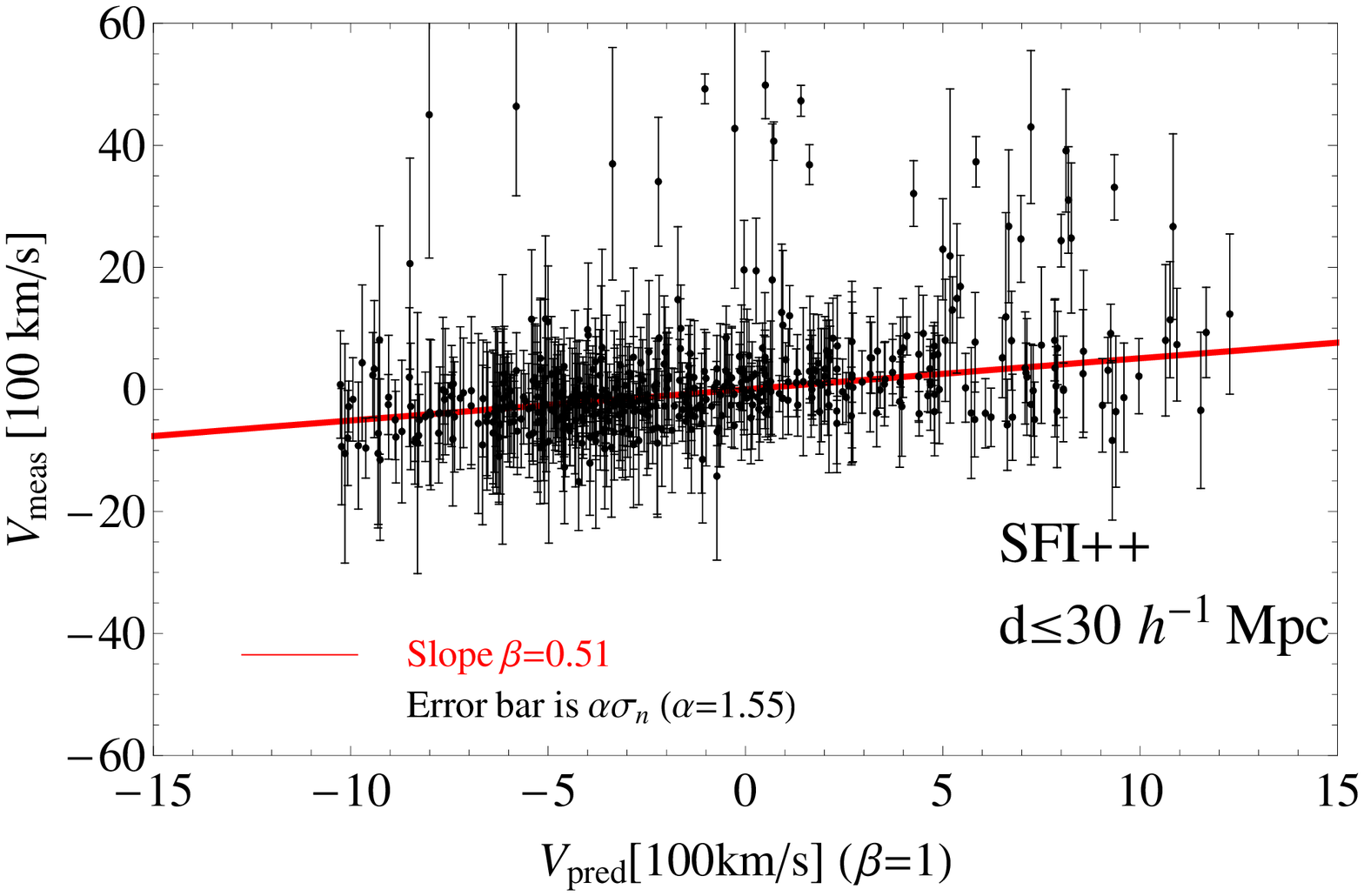}
\includegraphics[bb=0 0 524 343,width=3.0in]{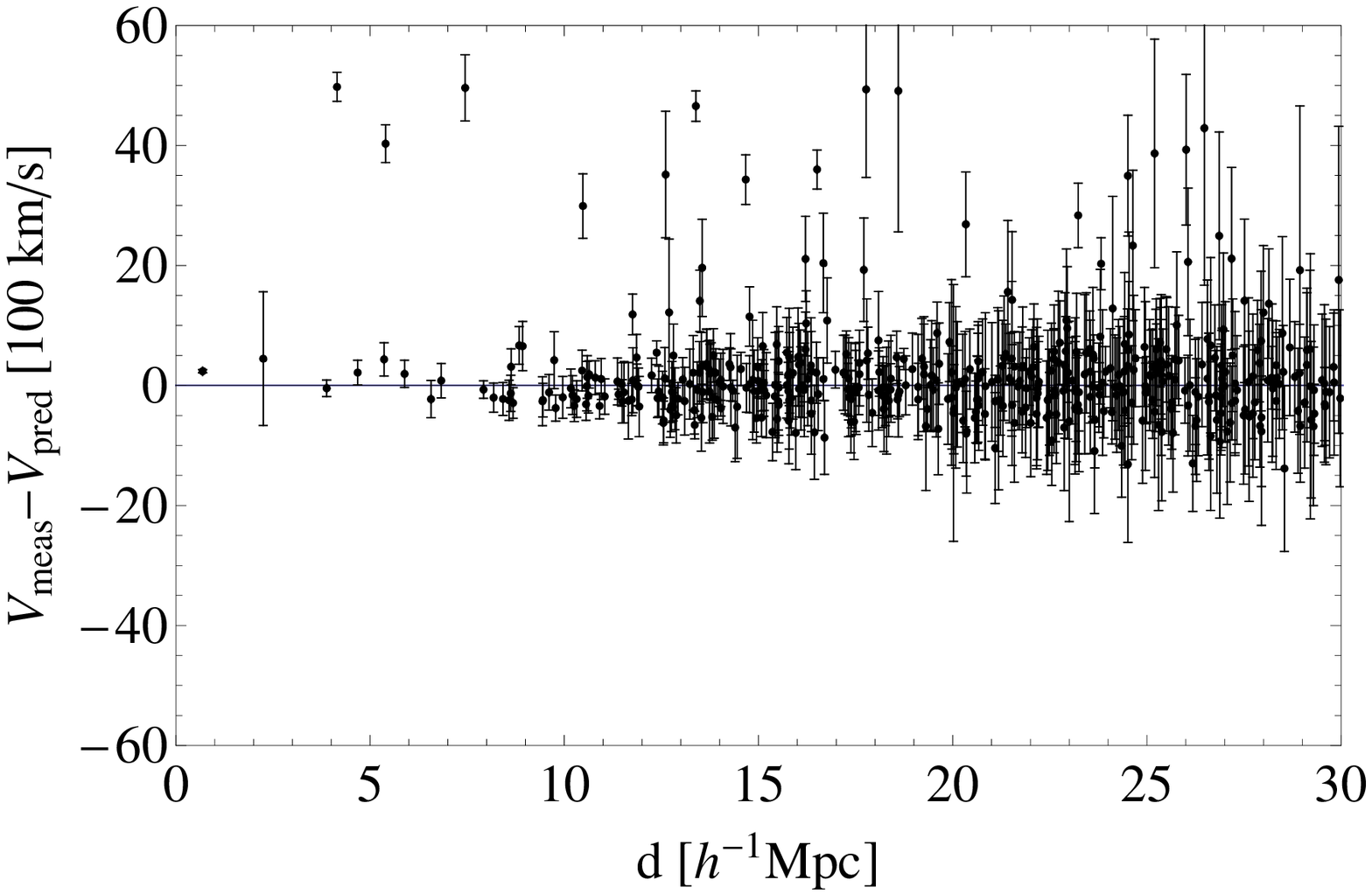}}
\centerline{\includegraphics[bb=0 0 643
420,width=3.0in]{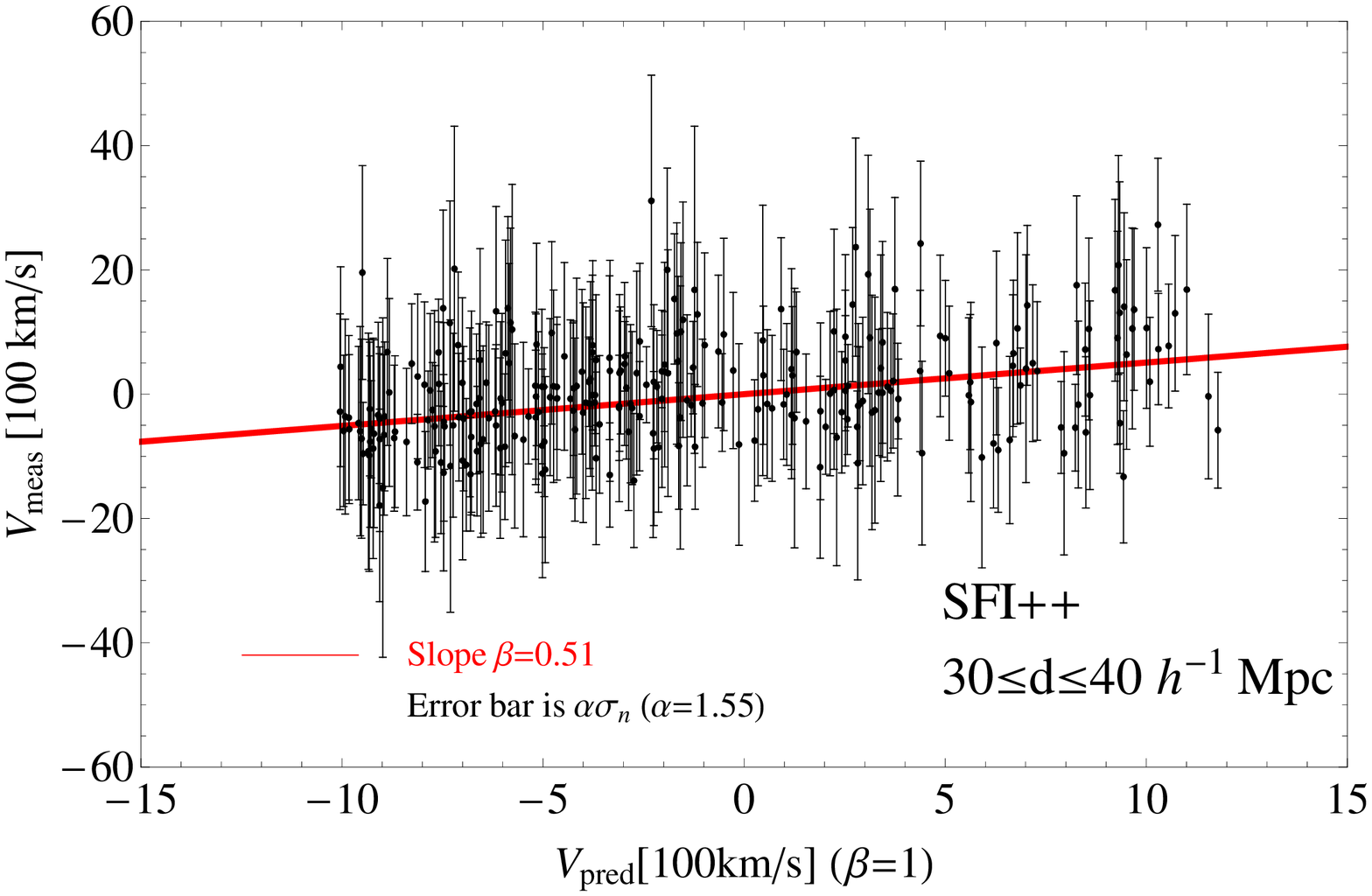}
\includegraphics[bb=0 0 546 356,width=3.0in]{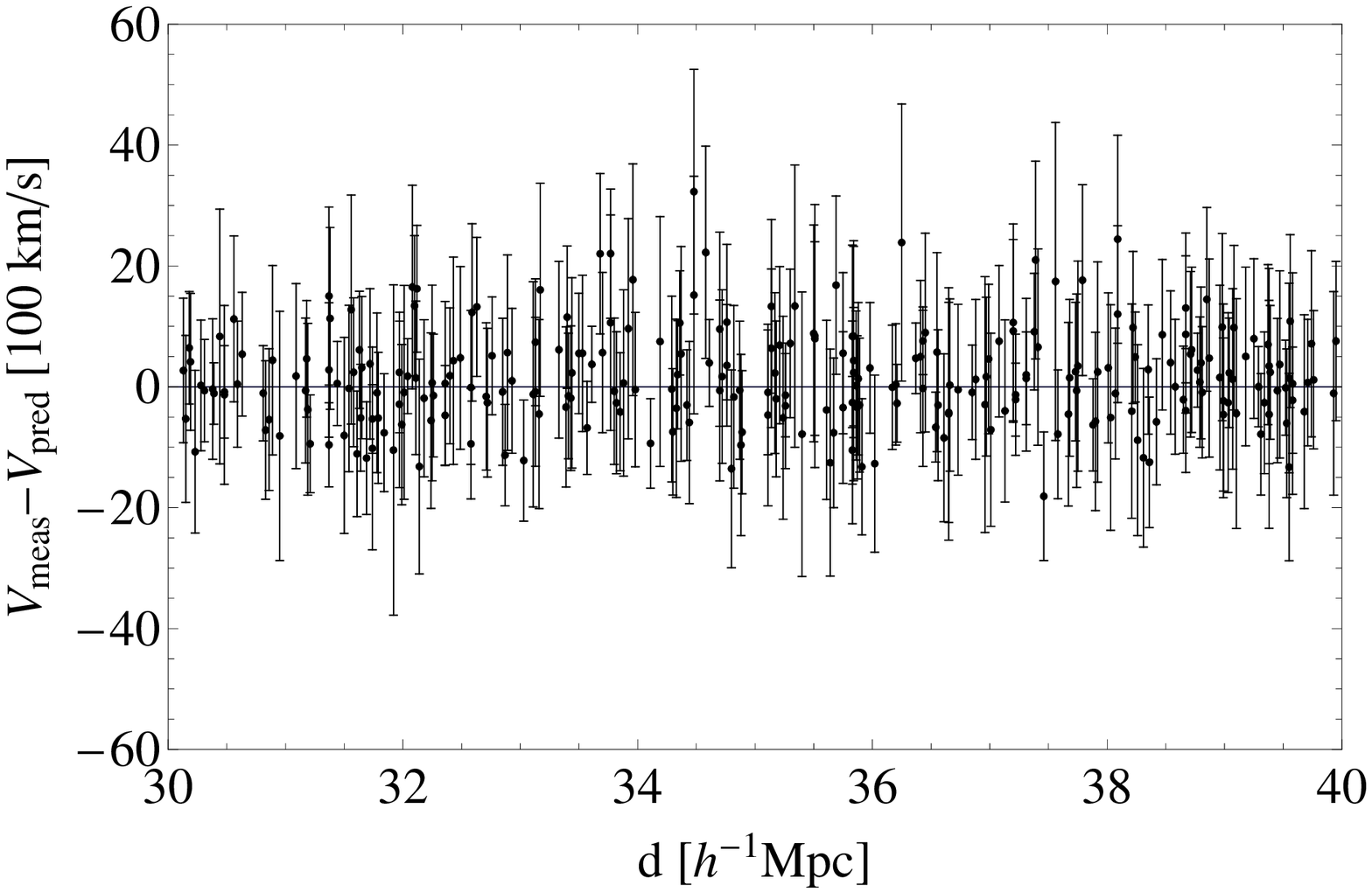}}
\centerline{\includegraphics[bb=0 0 663
435,width=3.0in]{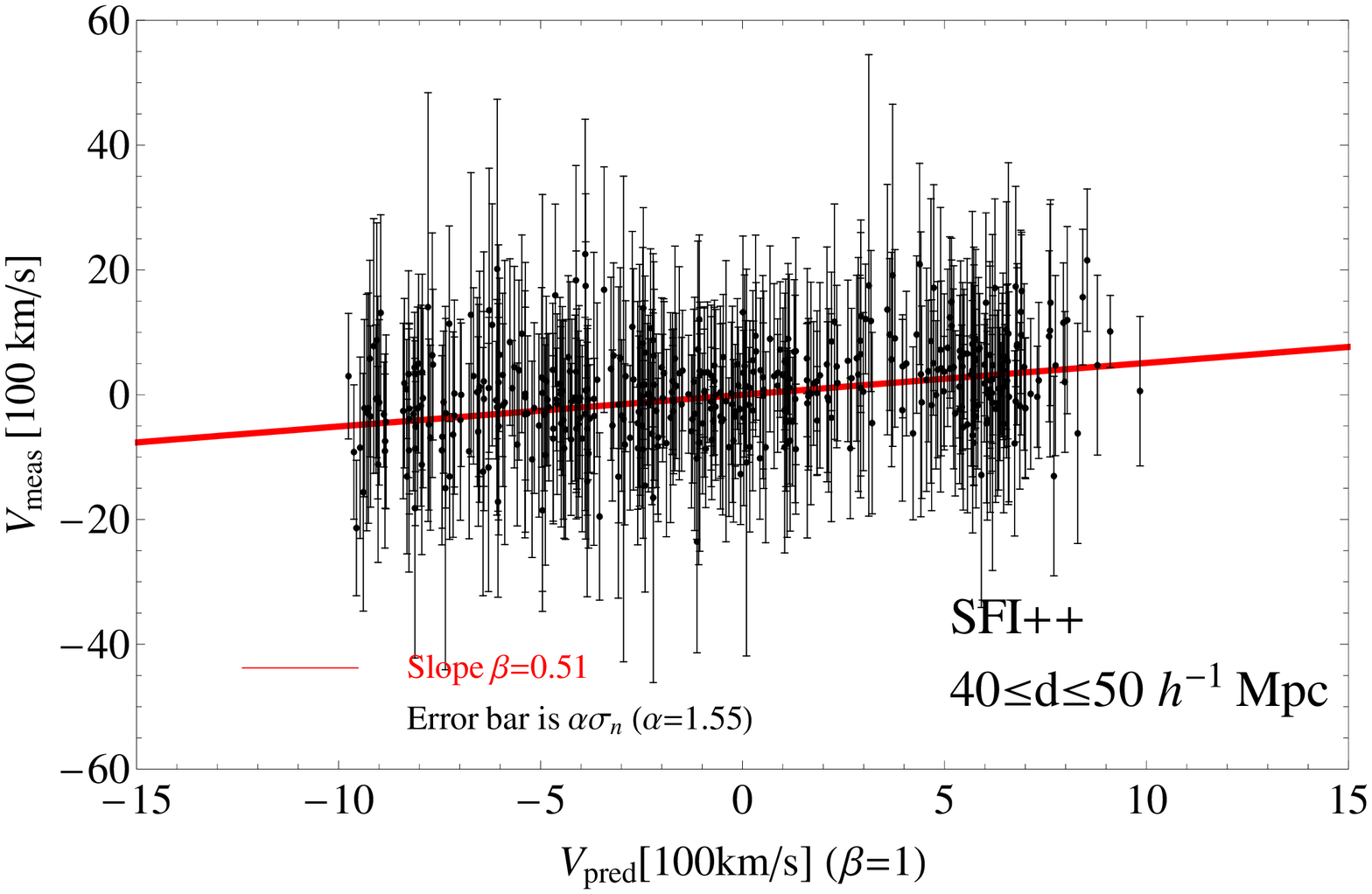}
\includegraphics[bb=0 0 542 353,width=3.0in]{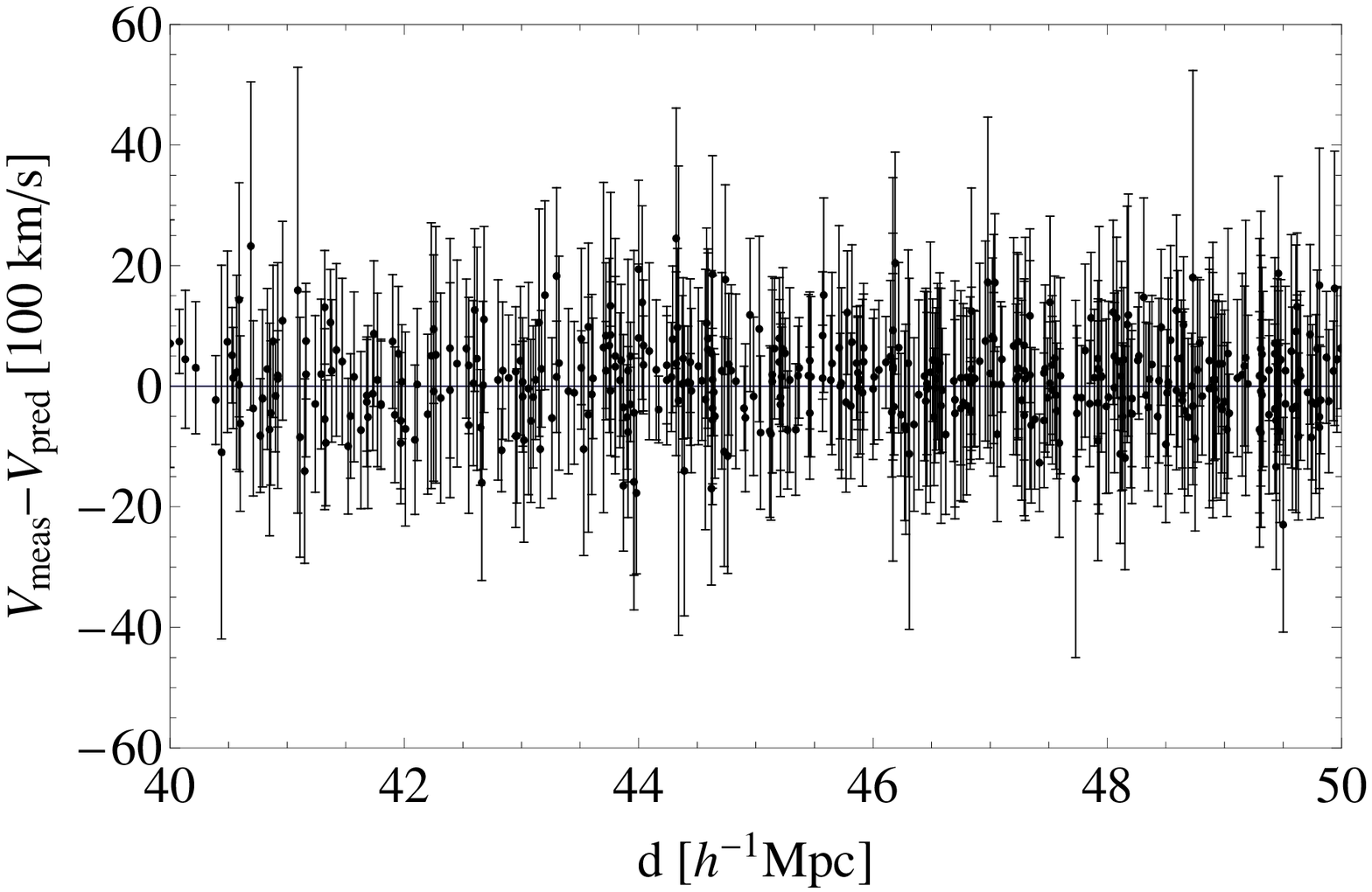}}
\centerline{\includegraphics[bb=0 0 758
494,width=3.0in]{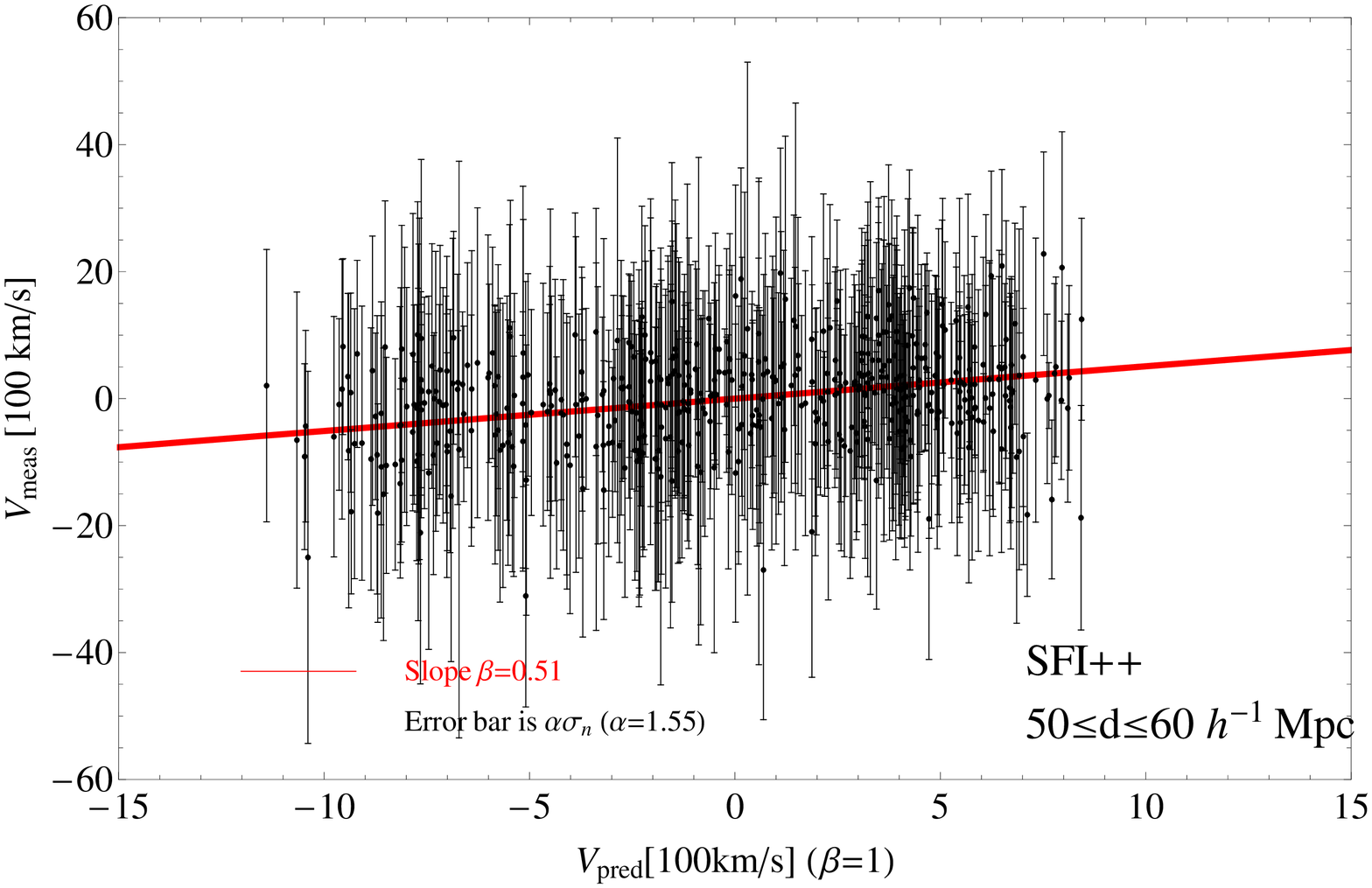}
\includegraphics[bb=0 0 557 362,width=3.0in]{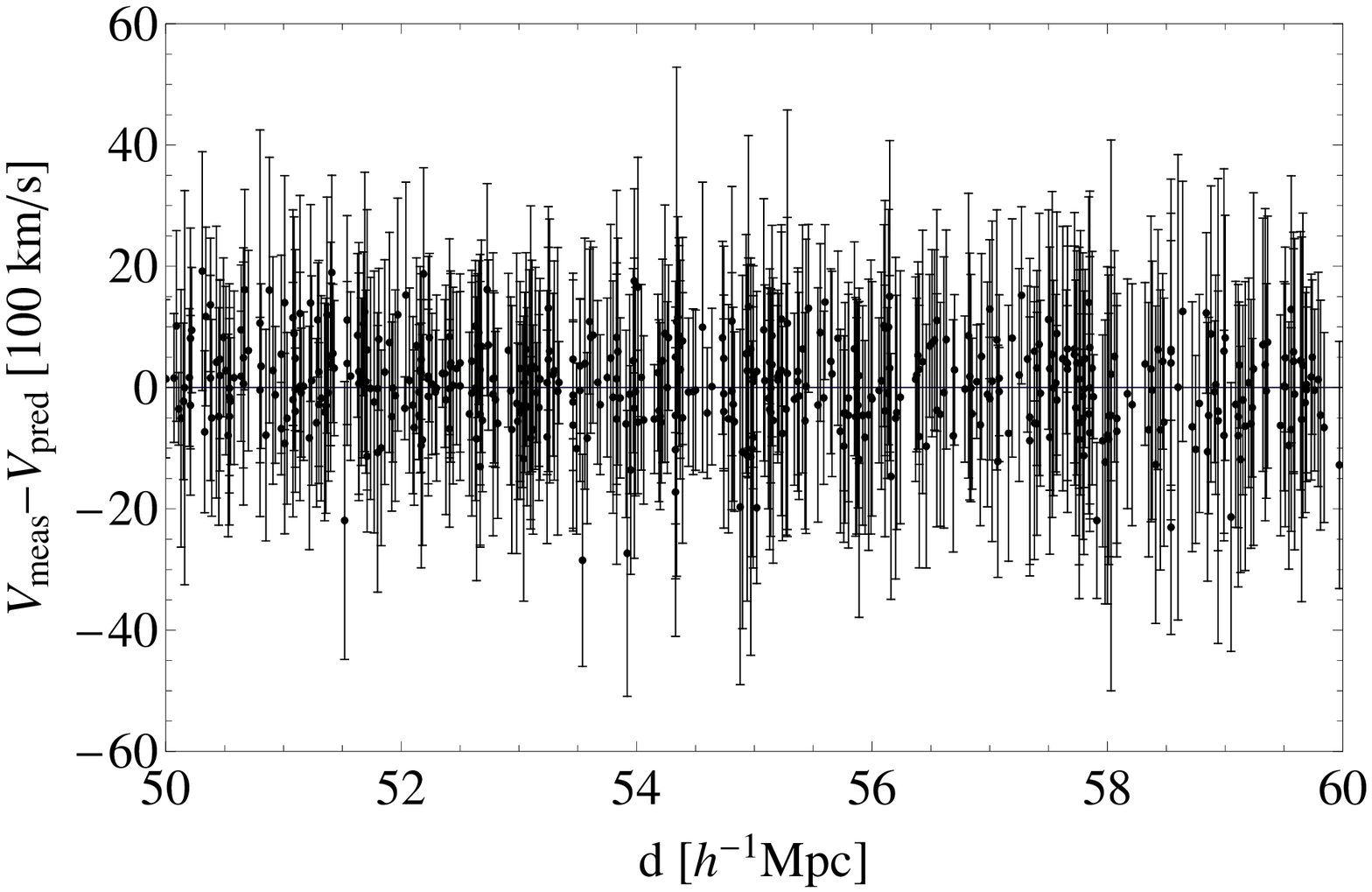}}
\caption{Same as Fig.~\ref{enearplot1} for the SFI++ catalogue.} \label{sfiplot1}
\end{figure*}

\begin{figure*}
\centerline{\includegraphics[bb=0 0 777
507,width=3.0in]{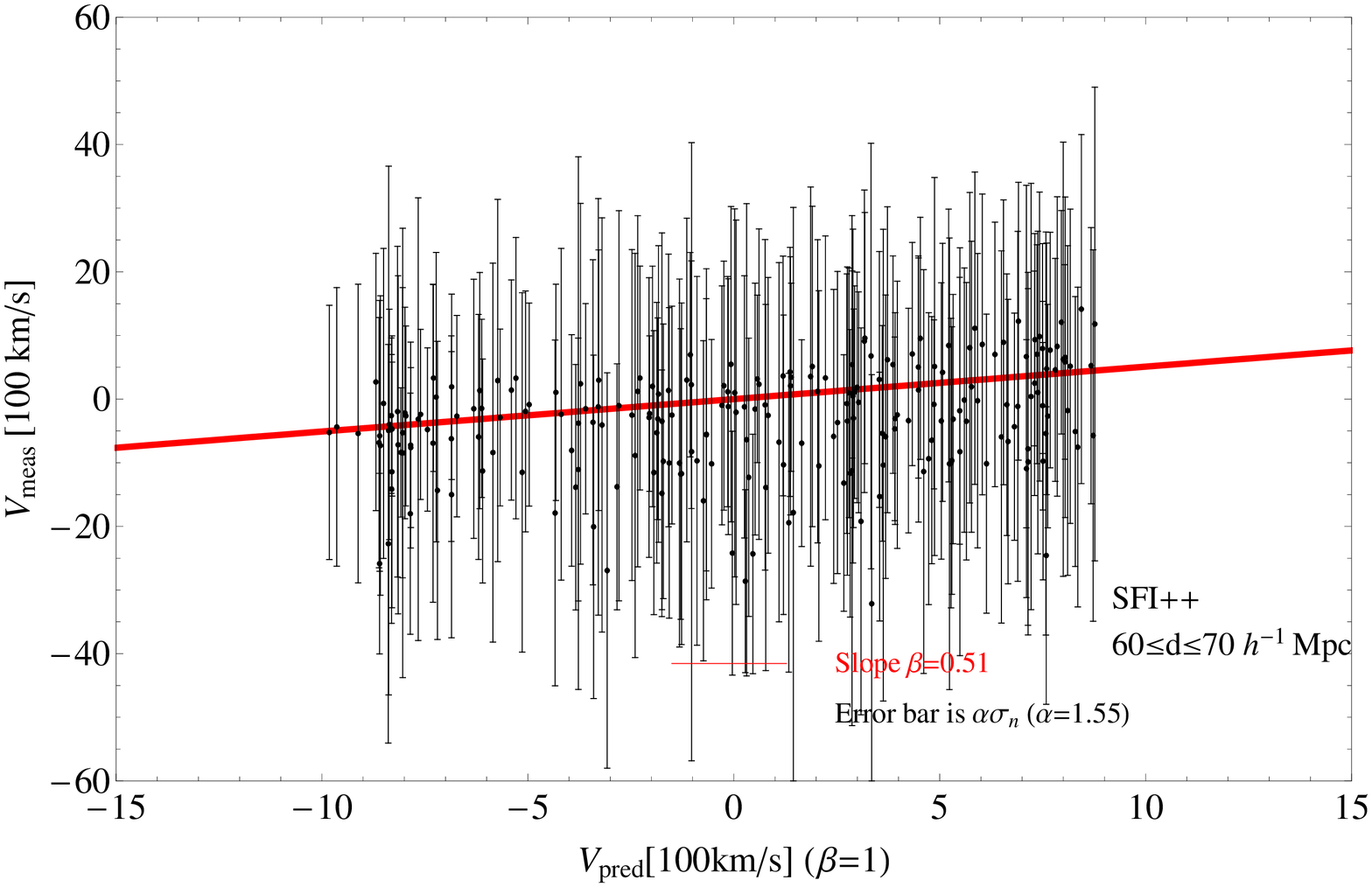}
\includegraphics[bb=0 0 574 374,width=3.0in]{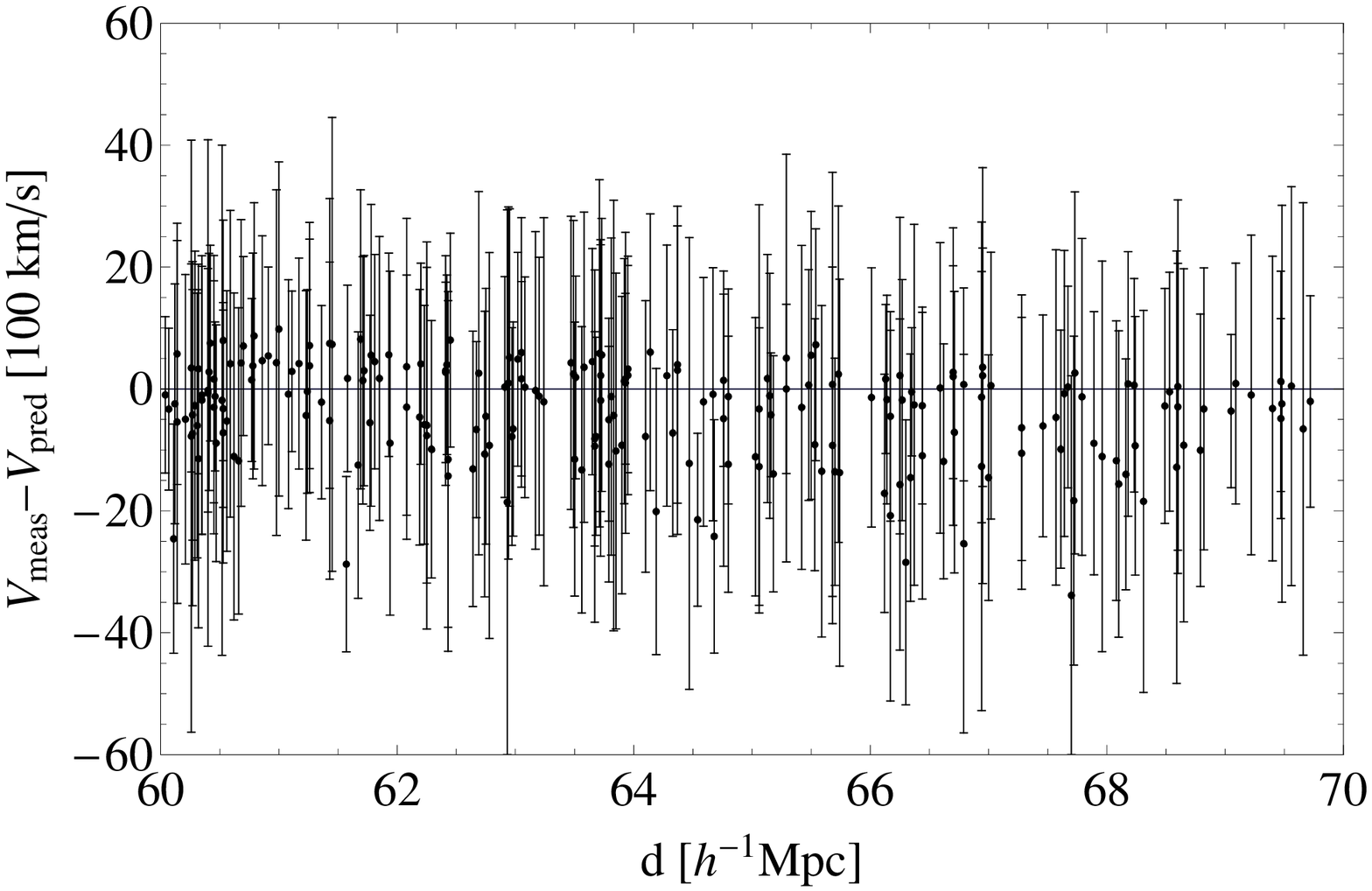}}
\caption{Same as Fig.~\ref{enearplot1} for the SFI++ catalogue $60
\leq d \leq 70 \hmpc$.} \label{sfiplot2}
\end{figure*}

Let us now investigate how well predicted peculiar velocities
match the observed ones in each catalogue. We do this by comparing
model predictions to observations on a point-by-point basis. The
results are the scatter plots shown in the left panels of Figs.
~\ref{enearplot1} to \ref{sfiplot2}. One point in each plot
indicates an object in one catalogue. Observed velocities are on
the
 $y$-axis and predicted velocities normalised to $\beta=1$
are indicated on the  $x$-axis.
 Error bars represent 1$\sigma$ uncertainties in the measured velocities multiplied
by the hyper-parameter of the catalogue, as indicated in
Table~\ref{tab2}. Straight lines represent the best fit value of
$\beta$ from the hyper-parameter analysis with a slope $\beta$
also listed in Table~\ref{tab2}. Panels on the right show the
velocity residuals
 $v_{\rm{meas}}-v_{\rm{rec}}(\beta=1)*\beta_{\textrm{best-fit}}$ computed at the estimated
 `true' distance of the object.
For those catalogues that contain a sufficiently large number of
objects, we break down the comparison by distance and show the
$v$--$v$ scatter plots for objects in different spherical shells.

\begin{enumerate}

\item ENEAR catalogue. In Fig.~\ref{enearplot1} we show  the
$v$--$v$ comparisons for objects in three different distance
intervals. In the innermost shell ($d<35 \hmpc$) the general
agreement between observed and predicted velocities is quite good.
The  few objects that show significant departures from the best
fit are typically found beyond $20 \hmpc$ and have positive
predicted velocities. Similar residuals are also seen in the
ENEAR-PSC$z$ velocity maps of  \cite{Nusser01} and are mostly
galaxies infalling in the Hydra-Centaurus direction and outflowing
from the Perseus-Pisces complex. The Hydra-Centaurus infall is
attributed to the Great Attractor, an overdensity originally
estimated to have a mass of $\sim 5 \times 10^{16} \rm{M}_{\odot}$
located at $(l,b,cz) \sim(307^{\circ},7^{\circ}, 4350 \hmpc)$
\citep{Lynden-Bell88a}, subsequently corrected to $M \sim 8 \times
10^{15} \rm{M}_{\odot}$  and ($289^{\circ},19^{\circ}, 3200
\hmpc$) \citep{Tonry00}. The absence of large velocity residuals
in this shell indicates that this prominent infall is well
reproduced by the model velocity field. This agreement persists
out to  $55 \hmpc$, i.e. in the second shell, meaning that the
possible backside infall to the Great Attractor is also correctly
predicted by the model.

Measured peculiar velocities in the outermost shell are, on
average, larger than model predictions. As anticipated, the
accuracy of the velocity model decreases with distance because
distant structures are poorly sampled by galaxies in the
flux-limited PSC$z$ catalogue. One example is the Shapley
concentration, a large complex of clusters whose dynamical
relevance has been outlined by many authors (e.g.
\citealt{Scaramella89,Hudson99b,Branchini99}). However, a poor
sampling of this supercluster would lead to an underestimate of
the predicted velocities, whereas here we have the opposite
effect. The systematic trend observed in the scatter plot would
rather be explained by the poor sampling of low density regions,
i.e. large super-voids. Alternatively, these negative residuals
could reflect the fact that our model velocity field relies on
linear theory and therefore tends to overestimate peculiar
velocities near the peaks of the density field, which in turn at
large distances could be artificially boosted up by  shot noise
\citep{Branchini00}. Whatever the reason, it is worth pointing out
that negative velocity residuals of the same amplitude were also
found by \cite{Nusser01} for ENEAR galaxies at similar distances
and concentrated in the area $l\sim 0^{\circ}$ and
$-60^{\circ}<b<-15^{\circ}$.

\item SN and A1SN catalogues. The $v$--$v$ comparisons of the
objects in both catalogues shown in Figs.~\ref{snplot1} and
\ref{a1snplot1} indicate a good match between model predictions
and observed peculiar velocities. This agreement is particularly
impressive considering the comparatively smaller errors in the
measured supernova velocities. This result highlights the fact
that Type Ia supernova samples are effective probes of the
underlying density field and should be considered as a  serious
alternative to galaxy-based, peculiar velocity catalogues.

 \item SFI++ catalogues.
 In this case we show the $v$--$v$ scatter plots for objects in
five different shells. Errors become progressively larger with
distance, having little statistical significance in the outer
shell.

In the innermost shells we notice the presence of several
discrepant data points, characterised by extremely large peculiar
velocities, all of them outgoing. No other object in any other
catalogues or in the more external shells has measured velocities
of the same magnitude. Note that for SFI++ galaxies we did not
perform our MB correction, but rely on the built-in MB
corrections. Therefore, the features of large velocities may
suggest an inadequate correction for the inhomogeneous MB. This is
quite challenging since in the very local region in which the
redshift surveys are used to trace the density field are often
incomplete.

Finally, in the outermost shells we see that residuals become
systematically more negative, in analogy to what we see in the
ENEAR catalogue. Explanations proposed to account for that
behavior are also valid here.

\end{enumerate}

\subsection{Analysis of the velocity residuals}
\begin{figure*}
\centerline{\includegraphics[bb=0 0 482
318,width=2.3in]{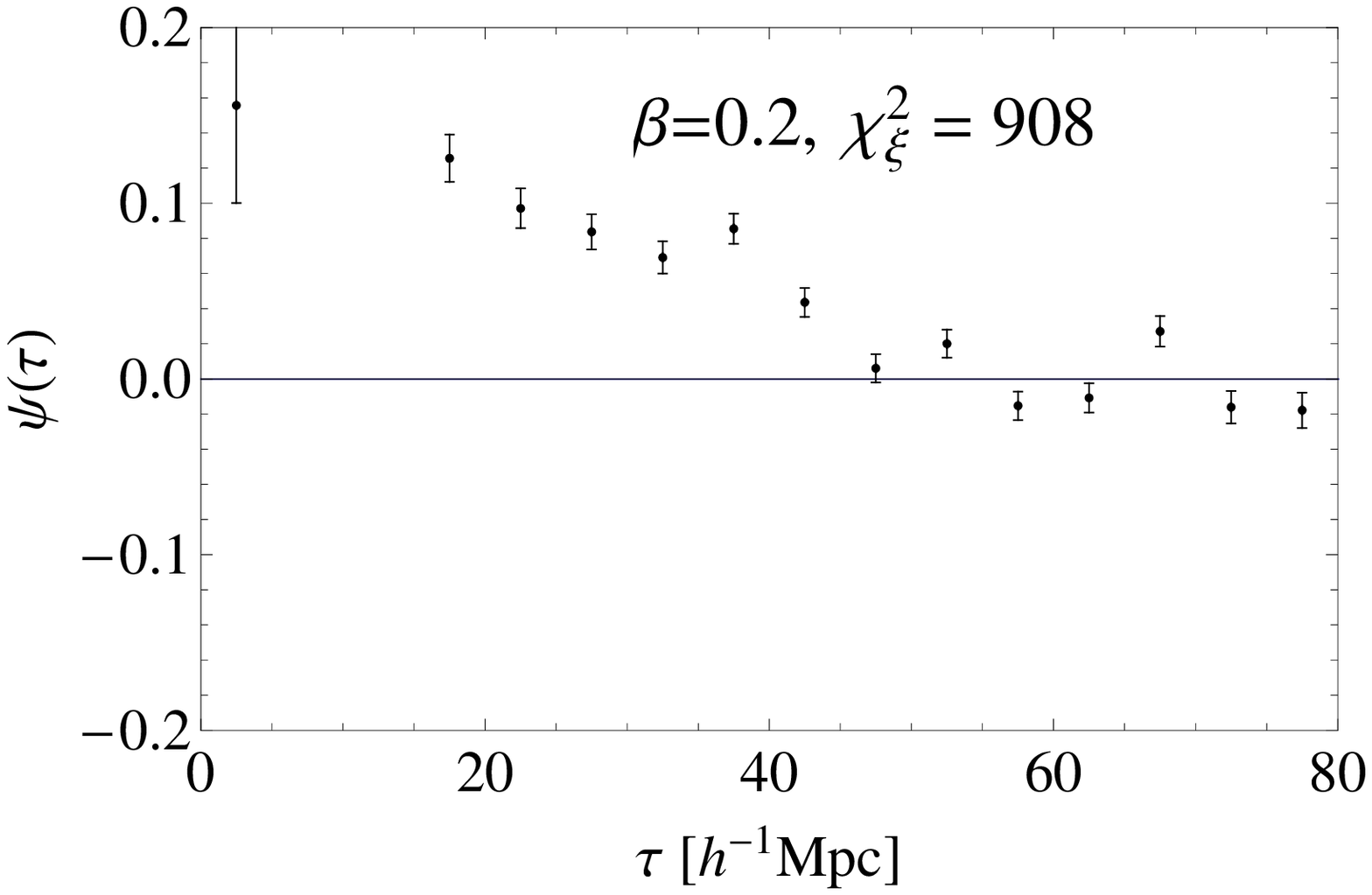}
\includegraphics[bb=0 0 494 322,width=2.3in]{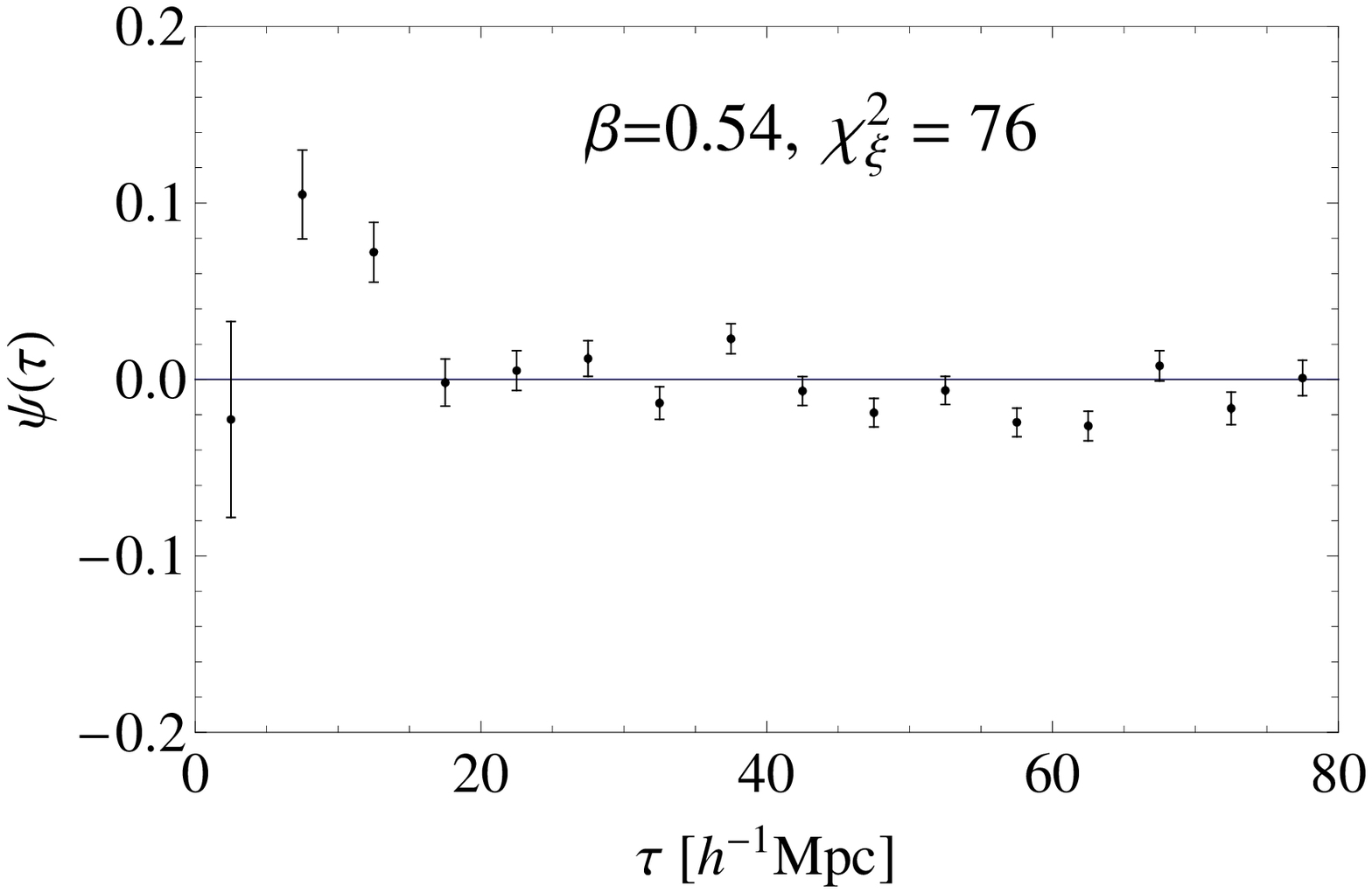}
\includegraphics[bb=0 0 506 332,width=2.3in]{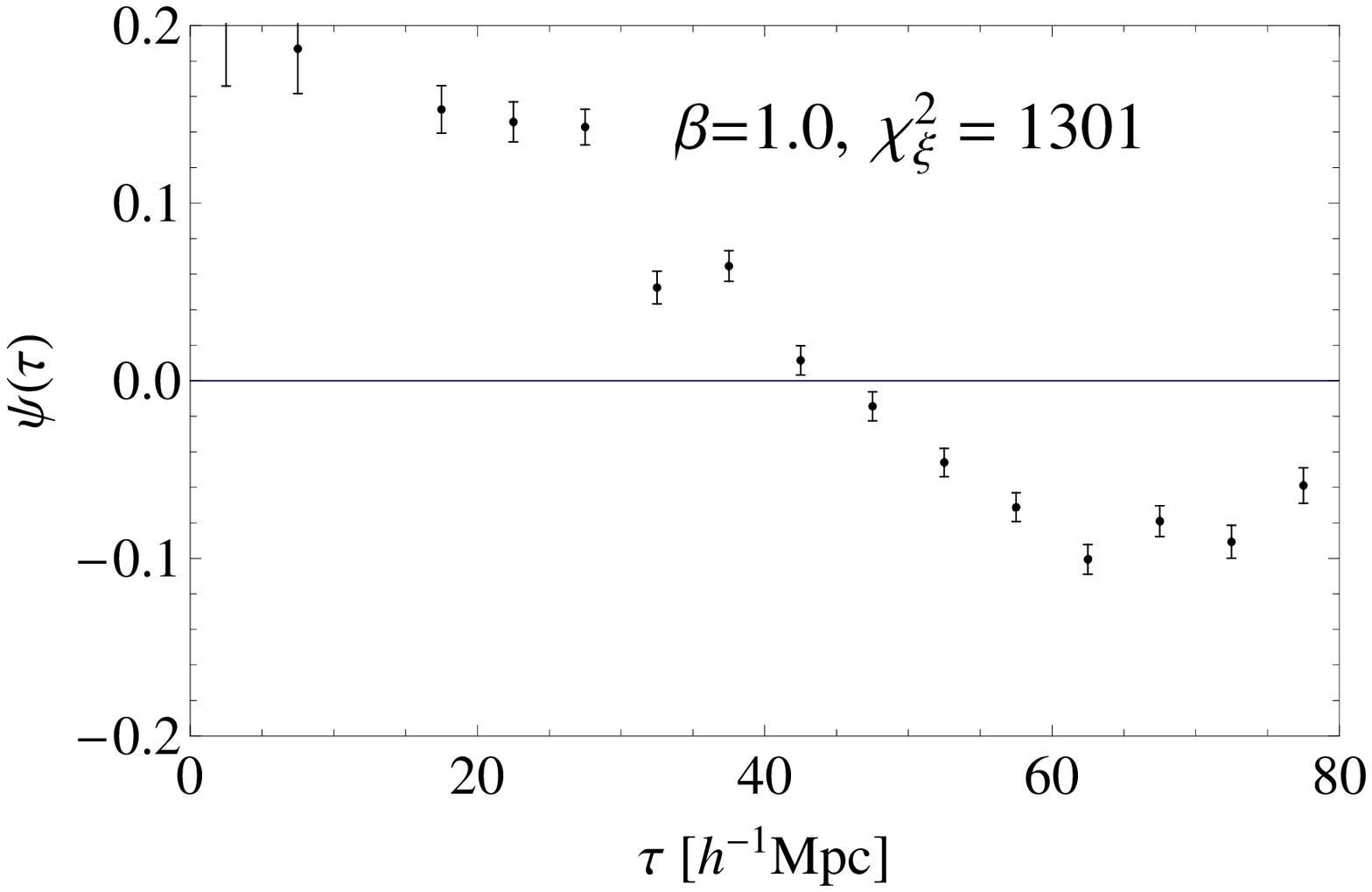}}
\caption{Correlation function of velocity residuals plotted for
$\beta=0.2$ (left), $0.54$ (middle) and $1.0$ (right). The samples
plotted are those pairs with $d_{\textrm{sep}} \leq 80 \hmpc$.}
\label{correlation-plot}
\end{figure*}

The Bayesian, hyper-parameter analysis determines the best $\beta$
value that characterises our model velocity field but cannot
address the question of whether or not the model provide an
adequate fit to the observed velocities. Inadequate fits typically
generate spurious correlations in the velocity residuals. Here we
follow \cite{Willick97, Willick98} and \cite{Branchini01} and look
for anomalous spatial correlation in the $v$--$v$ residual maps.

First, we  define the normalised velocity residual as
\begin{equation}
\delta_{i}(\beta)=\frac{v_{\textrm{meas},i}-\beta*v_{\textrm{rec},i}(\beta=1)}{\sigma^{\textrm{mea}}_{i}},
\end{equation}
for each galaxy in the catalogue. This quantity depends on the
free parameter $\beta$ and, when averaged over all objects in the
catalogues, is minimised by the corresponding best fit values in
Table~\ref{tab2}. Then, we consider all pairs of galaxies in the
catalogue and compute the two-point correlation function for the
velocity residuals:
\begin{equation}
\psi(\tau)=\frac{1}{N(\tau)}\sum_{i<j}\delta_{i}\delta_{j},
\label{correlate1}
\end{equation}
where $N(\tau)$ is the number of galaxy pairs within predicted
separation $d_{i,j} \leq \tau \pm 5 \hmpc$ and the sum runs over
all pairs $(i,j)$. If the model velocity field provides a good
match with the observed velocities then residuals should be
spatially uncorrelated. In contrast, correlated residuals indicate
the presence of systematic effects on some scale.

Normalised velocity residuals and their correlation function have
been computed for all objects in the catalogues and for values of
$\beta$ in the range $[0.1,1]$ in steps of $\Delta\beta=0.1$. As
an example in Fig.~\ref{correlation-plot} we show the correlation
function of the velocity residuals, $\psi(\tau)$, for all ENEAR
galaxies with separations $d_{\rm{sep}}<80 \hmpc$. The three
panels refer to different values of $\beta$ specified by the
labels, including the best-fit value (central panel). Error bars
represent Poisson noise, $N(\tau)^{-0.5}$. It is clear form the
plots that the best fit model behaves much better that the two
other cases explored, in which residuals are significantly
correlated (or anti-correlated) at almost all separations.
Focusing on the $\beta=0.54$ case we see that the value of
$\psi(\tau)$ is consistent with the null hypothesis of no
correlation over most of the distance bins. The excess correlation
at small scales reflects the fact that the model velocity field is
smoothed with a Gaussian filter of radius $5 \hmpc$. The residual
correlation functions of all other catalogues are similar to those
of the ENEAR galaxies and therefore are not shown here.

To quantify the goodness of fits we compute the following quantity
\citep{Branchini01}
\begin{equation}
\chi^{2}_{\xi}=\sum^{N_{\textrm{bins}}}_{k=1}\frac{\xi^{2}(\tau_{k})}{N(\tau_{k})},
\label{chi2xi}
\end{equation}
where $\xi(\tau)=N(\tau)\psi(\tau)$, and $N_{\textrm{bins}}$ is
the number of  bins in which we compute  $\psi(\tau)$.
 \cite{Willick97} showed that if residuals are uncorrelated
then $\xi(\tau)$ behaves like a Gaussian random variable with zero
mean and variable $N(\tau)$, and that the $\chi^{2}_{\xi}$ is
distributed like a  $\chi^2$ function with a number of degrees of
freedom equal to $N_{\rm eff}\sim0.87\times N_{\rm bins}$, where
the factor $0.87$, computed from mock catalogues,  accounts for
the correlation among the bins.

We have computed  $\chi^{2}_{\xi}$  as a function of $\beta$  for
all velocity catalogues. Results are shown in panel (c) of
Fig.~\ref{like-plot}. Different symbols and colours represent
different velocity catalogues, indicated by the labels. To ease
the comparison we set the minimum of $\chi^{2}_{\xi}$ equal to
zero for all curves. The $\chi^2$ distributions are remarkably
 similar, with minima very close to each other and in agreement with the
 best fit value of $\beta$. This result is by no means trivial and indicate
 that the PSC$z$ linear velocity model with $\beta=0.53$ provides an adequate
 fit to the peculiar velocities of very different types of objects from
 different  catalogues.
To quantify the agreement we exploit the fact that
$\chi^{2}_{\xi}$ obeys $\chi^{2}$ statistics and use this fact to
compute a formal 1$\sigma$ error from $\Delta \chi^{2}_{\xi}=1$.
The results of this exercise are listed in column 3 of
Table~\ref{tab2} and confirm the qualitative agreement found from
the visual inspection.

\section{Discussion and Conclusion}
\label{sec-discuss}


In this work we used the PSC$z$ galaxy redshift catalogue to
predict the cosmic velocity field in the local Universe within the
framework of gravitational instability  and assuming
 linear theory and linear biasing. The model velocity field, which depends on
 a single parameter $\beta$, is
compared to the observed velocities of different types of objects
 in the ENEAR, SN, SFI++ and A1SN catalogues.
We restrict our comparison to objects within $70 \hmpc$, where
errors on measured velocities are reasonably small and we can
trust the model velocity field. Great care has been taken in
correcting for the inhomogeneous Malmquist bias. This is done by
using PSC$z$ to trace the underlying mass density field and by
adopting a Monte Carlo rejection procedure to statistically
correct for the MB. Finally, model and observed velocities have
been compared in a Bayesian framework using the hyper-parameter
method. This technique is designed to estimate the statistical
weights  of different data set in an objective way and allows one
to jointly analyse different velocity catalogues. Here are the
main results of our analysis.

\begin{enumerate}

\item The hyper-parameter $v$--$v$ comparisons performed for each
catalogue give $\beta$ values that are consistent within the
errors. This means that the single velocity model is a best fit to
all data sets. In addition, the best fit  $\beta$ values for the
ENEAR, SN and A1SN catalogues
 agree with those obtained from previous $v$--$v$ comparisons that used the
 same  data sets but  different velocity models and comparison techniques.

\item The joint $v$--$v$ comparison performed with the
hyper-parameter technique using all data sets significantly
improves the accuracy in the estimate of $\beta$. The best fit
value is $\beta=0.53\pm0.014$.

This result can be used to set a constraint on the growth rate of
density fluctuations at $z\sim0$, i.e. $f_{0}$. The value of this
quantity is sensitive to the expansion history of the Universe.
Its measurement will allow us to understand whether the current
cosmic acceleration is driven by a Dark Energy component or
perhaps general relativity theory needs modification on
cosmological scales. Most of the current estimates have been
performed at moderate redshift by quantifying the apparent
anisotropies in the clustering of galaxies induced by redshift
space distortions (RSD,
\citealt{Percival04,Tegmark06,Guzzo08,Song09,Blake11,Reid12,Samushia12,Tojeiro12}).
Indeed one of the main goals of ongoing (\citealt{wigglez,boss};
Guzzo et al., in preparation) and planned \citep{bigboss, euclid}
redshift surveys is to measure the growth rate at higher redshift,
to increase the $z$-baseline for this cosmological test. As
pointed out for example by
 \cite{Hudson12}, the additional estimates at $z\sim 0$ can
 sharpen the observational constraints. Moreover, at $z\sim0$ the growth rate
 can be estimated using techniques alternative to RSD, hence providing an
 important cross-check among the methods.

In Table~\ref{tab3} we list the most recent estimate of
$f\sigma_8$, the growth rate normalised to the rms mass density
fluctuation on $8 \hmpc$ scales. The last entry is the value
obtained from our analysis computed as
$f\sigma_8=\beta\sigma^{\textrm{gal}}_{8} $, where
$\sigma^{\textrm{gal}}_{8} \simeq 0.80 \pm 0.05$ is the rms
fluctuation in the number density of PSC$z$ galaxies
\citep{Hamilton02}. This result agrees with  $\Lambda$CDM model
predictions. All estimates in Table~\ref{tab3} are in reasonably
good agreement. The largest discrepancy is below 2$\sigma$
significance.

\begin{table}
\begin{centering}
\begin{tabular}{@{}lcl}\hline
Comparison & $f\sigma_{8}$  & Reference
\\ SFI++ \textit{vs} 2MRS & $0.31 \pm 0.04$ & \cite{Davis11}
\\ SN \textit{vs} PSC$z$ & $0.44 \pm 0.06$ & \cite{Radburn-Smith04}
\\ A1SN \textit{vs} PSC$z$ & $0.40 \pm 0.07$ & \cite{Turnbull12}
\\ 6dF (RSD) & $0.42 \pm 0.05$ & \cite{Beutler12}
\\  \textbf{Combined} \textit{vs} \textbf{PSC$z$}  & \textbf{$0.42 \pm
0.033$} & \textbf{This study}
\\\hline
\end{tabular}%
\caption{Constraints on $f\sigma_{8}$ from various catalogues. In
the $\Lambda$CDM model, $f\simeq \Omega^{0.55}_{\rm{m}}$
\citep{Peebles71,Linder05}.} \label{tab3}
\end{centering}
\end{table}

\item Hyper-parameters. From the hyper-parameter analyses of the
$v$--$v$ comparison and after marginalising over $\beta$ we obtain
the hyper-parameters of the different data sets. We find that all
of them are slightly larger but quite close to unity, which
indicates that velocity errors are slightly overestimated in all
velocity catalogues, possibly reflecting the fact that in our
analysis we did not include uncertainties in the MB correction.

\item $v$--$v$ scatter plots. The inspection of the $v$--$v$
comparison on a point-by-point basis from scatter plots reveals
that, on average,  the model velocity field provides a good fit to
observed peculiar velocities in all catalogues. The agreement is
especially remarkable for the case of Type Ia supernovae that have
much smaller velocity errors and constitutes an important check
for the gravitational instability scenario.

The only exceptions are distant ENEAR and SFI++ galaxies, for
which model prediction systematically overestimate observed
velocities. A similar systematic was  seen in a previous ENEAR vs.
PSC$z$ comparison \citep{Nusser01}, mainly for galaxies in the
region $l\sim 0^{\circ}$ and $-60^{\circ}<b<-15^{\circ}$. Among
the possible explanations is the possible poor sampling of distant
inhomogeneities or inaccurate modelling of the flow around distant
density peaks. Additionally, a few nearby SFI++ galaxies have
peculiar velocities that are very large and  cannot be matched by
model predictions. This might just reflect uncertainties in the
original correction for inhomogeneous Malmquist bias which, in the
local volume, is notoriously difficult to model due to the
incompleteness in the parent redshift catalogues.

\item Goodness of the fit. To evaluate whether our best fit model
velocity field is also a {\it good} model we searched for spurious
correlations among velocity residuals. We have considered all
object pairs with separations up to $80 \hmpc$. Apart from a
positive correlation signal at small separations, induced by the
Gaussian window used to filter out non-linear effects, when we set
$\beta$ equal to its best fit value we find that the residual
correlation
 function is consistent with zero. This result further confirms the adequacy of the
 velocity model and the validity of the gravitational instability picture.

\end{enumerate}

Although the study of the peculiar velocity field is still very
data-limited, future surveys, e.g. the 6dF survey \citep{Jones09},
or eventually using the 21 cm line, e.g. with the \cite{ska}, may
provide rich resources for the study of large-scale structure and
cosmic flows. But even with these large data-sets, it will still
be important to carefully assess the potential systematics.

\vskip 0.1 truein

\noindent \textbf{Acknowledgments:} We would like to thank George
Efstathiou and Jeremiah Ostriker for helpful discussions, and
Michael Hudson and Stephen Turnbull for sharing A1SN catalogue.
YZM is supported by a CITA National Fellowship. This research is
supported by the Natural Science and Engineering Research Council
of Canada.



\begin{thebibliography}{}
\bibitem[\protect\citeauthoryear{Bernardi et al.}{2002}]{Bernardi02} Bernardi M.,
Alonso M. V., da Costa L. N., Willmer C. N. A., Wegner G.,
Pellegrini P. S., Rite C., Maia M. A. G., 2002, AJ, 123, 2990



\bibitem[\protect\citeauthoryear{Beutler et al.} {2012}]{Beutler12} Beutler F., et al.,
2012, MNRAS, 423, 3430

\bibitem[\protect\citeauthoryear{Blake et al.} {2011}]{Blake11} Blake C., et al.,
2011, MNRAS, 415, 2876

\bibitem[\protect\citeauthoryear{Branchini et al.} {2001}]{Branchini01} Branchini E., 2001,
MNRAS, 326, 1191

\bibitem[\protect\citeauthoryear{Branchini et al.} {1999}]{Branchini99} Branchini E., et al.,
1999, MNRAS, 308, 1

\bibitem[\protect\citeauthoryear{Branchini et al.} {2000}]{Branchini00} Branchini
E., Zehavi I., Plionis M., Dekel A., 2000, MNRAS, 313, 491

\bibitem[\protect\citeauthoryear{Branchini et al.} {2012}]{Branchini12} Branchini E., Davis M., Nusser A.,
2012, MNRAS, 424, 472

\bibitem[\protect\citeauthoryear{Colless et al.} {2001}]{Colless01} Colless M., Saglia R. P.,
Burstein D., Davis R. L., McMahan R. K., Wegner G., 2001, MNRAS,
321, 277

\bibitem[\protect\citeauthoryear{da Costa et al.} {1998}]{Costa98} da Costa L. N., Nusser A., Freudling W.,
Giovanelli R., Haynes M. P., Salzer J. J., Wegner G., 1998, MNRAS,
299, 425

\bibitem[\protect\citeauthoryear{da Costa et al.} {2000}]{Costa00} da Costa L. N., Bernardi M.,
Alonso M. V., Wegner G., Willmer, C. N. A., Pellegrini P. S.,
Rit¨¦ C., Maia M. A. G., 2000, AJ, 120, 95

\bibitem[\protect\citeauthoryear{Dale et al.} {1999}]{Dale99} Dale D. A.,
Giovanelli R., Haynes M. P., Campusano L. E., Hardy E., 1999, AJ,
118, 1489

\bibitem[\protect\citeauthoryear{Davis et al.} {1996}]{Davis96} Davis M., Nusser A.; Willick J. A., 1996, ApJ,
473, 22

\bibitem[\protect\citeauthoryear{Davis et al.} {2011}]{Davis11} Davis M., Nusser A., Masters K. L.,
Springob C., Huchra J. P., Lemson G., 2011, MNRAS, 413, 2906

%

\bibitem[\protect\citeauthoryear{Drinkwater et al.}{2010}]{wigglez} Drinkwater M. J., et al., 2010, MNRAS, 401, 1429

\bibitem[\protect\citeauthoryear{Eisenstein et al.}{2011}]{boss} Eisenstein D. J., et al., 2011, AJ, 142, 72

\bibitem[\protect\citeauthoryear{Folatelli et al.} {2010}]{Folatelli10} Folatelli G. et al.,
2010, AJ, 139, 120

\bibitem[\protect\citeauthoryear{Feldman et al.} {2001}]{Feldman01} Feldman H. A., Frieman J. A.,
Fry J. N., Scoccimarro R., 2001, Phys. Rev. Lett., 86, 1434


\bibitem[\protect\citeauthoryear{Feldman et al.} {2010}]{Feldman10} Feldman H., Watkins R., Hudson M.
J., 2010, MNRAS, 407, 2328

\bibitem[\protect\citeauthoryear{Fisher et al.} {1995}]{Fisher1995} Fisher K., Huchra J., Strauss M.,
Davis M., Yahil A., Schlegel D., 1995, ApJ, 100, 69

%
\bibitem[\protect\citeauthoryear{Giovanelli et al.} {1997}]{Giovanelli97} Giovanelli R.,
Haynes M. P., Herter T., Vogt  N. P., Wegner G., Salzer J. J., da
Costa  L. N., Freudling W., 1997, AJ, 113, 22

\bibitem[\protect\citeauthoryear{Giovanelli et al.} {1998}]{Giovanelli98} Giovanelli R.,
Haynes M. P., Salzer J. J., Wegner G., da Costa L. N., Freudling
W., 1998, AJ, 116, 2632

\bibitem[\protect\citeauthoryear{Guzzo et al.} {2008}]{Guzzo08} Guzzo L., et al.,
2008, Nat, 541, 541


\bibitem[\protect\citeauthoryear{Hamilton and Tegmark} {2002}]{Hamilton02} Hamilton A.J.S.,
and Tegmark M., 2002, MNRAS, 330, 506

\bibitem[\protect\citeauthoryear{Haynes et al.} {1999}]{Haynes99} Haynes M. et al., 1999, Astron. J., 117, 2039


\bibitem[\protect\citeauthoryear{Hicken et al.} {2009}]{Hicken09} Hicken M. et al., 2009, ApJ, 700, 1097

\bibitem[\protect\citeauthoryear{Hobson et al.} {2002}]{Hobson02} Hobson M.P., Bridle S.L., Lahav O.,
2002, MNRAS, 335, 377


\bibitem[\protect\citeauthoryear{Huchra et al.} {2012}]{Huchra12} Huchra J.P., et al., 2012, ApJS, 199, 26



\bibitem[\protect\citeauthoryear{Hudson} {1994}]{Hudson94} Hudson M. J., 1994, MNRAS, 266,
468


\bibitem[\protect\citeauthoryear{Hudson} {1999a}]{Hudson99a} Hudson M. J., 1999a, PASP, 111, 57

\bibitem[\protect\citeauthoryear{Hudson} {1999b}]{Hudson99b} Hudson M. J., 1999b, ApJL, 512,
59

\bibitem[\protect\citeauthoryear{Hudson et al.} {2004}]{Hudson04} Hudson M.J., Smith R.J.,Lucey J. R.,
Branchini E., 2004, MNRAS, 352, 61

\bibitem[\protect\citeauthoryear{Hudson \& Turnbull} {2012}]{Hudson12} Hudson M.J.,
Turnbull S., 2012, ApJ, 751L, 30

\bibitem[\protect\citeauthoryear{Jha et al.} {2007}]{Jha07} Jha S., Riess A. G., Kirshner R. P., 2007, ApJ, 659, 122

\bibitem[\protect\citeauthoryear{Jones et al.} {2009}]{Jones09} Jones D. H., et. al, 2009, MNRAS, 399, 683

%


\bibitem[\protect\citeauthoryear{Kaiser} {1989}]{Kaiser89} Kaiser N., Lahav, O.,1989, MNRAS, 237, 129

%
%


\bibitem[\protect\citeauthoryear{Lahav et al.} {2000}]{Lahav00} Lahav O., Bridle S. L., Hobson M. P., Lasenby A. N., Sodre L.,
2000, MNRAS, 315, 45

\bibitem[\protect\citeauthoryear{Laureijs et al.}{2011}]{euclid} Laureijs R., et al., 2011, arXiv:1110.3193

\bibitem[\protect\citeauthoryear{Linder} {2005}]{Linder05} Linder E. V., 2005, Phys. Rev. D.,
72, 043529



\bibitem[\protect\citeauthoryear{Lynden-Bell et al.} {1988}]{Lynden-Bell88a} Lynden-Bell D.,
Faber S. M., Burstein D., Davis R. L., Dressler A., Terlevich
R.J., Wegner G., 1988, ApJ, 326, 19


\bibitem[\protect\citeauthoryear{Ma et al.} {2010}]{Ma10} Ma Y. Z., Zhao W., Brown M. L., 2010, JCAP,
1010, 007

%
\bibitem[\protect\citeauthoryear{Ma et al.} {2011}]{Ma11b} Ma Y.Z., Ostriker J., Zhao G.B., 2012, JCAP, 6, 26

\bibitem[\protect\citeauthoryear{Malmquist} {1920}]{Malmquist20} Malmquist K. G., 1920 Medd. Lund. Astron. Obs., Ser II,
22, 1


\bibitem[\protect\citeauthoryear{Masters et al.} {2006}]{Masters06} Masters K. L.,
Springob C. M., Haynes M. P., Giovanelli R., 2006, ApJ, 653, 861



\bibitem[\protect\citeauthoryear{Nusser et al.} {2001}]{Nusser01} Nusser A.,
da Costa L.  N., Branchini E., Bernardi  M., Alonso M. V., Wegner
G., Willmer C. N. A., Pellegrini P. S., 2001, MNRAS, 320, L21


\bibitem[\protect\citeauthoryear{Nusser et al.} {2011}]{Nusser11} Nusser A., Branchini E., Davis M., 2011, ApJ, 735, 77


\bibitem[\protect\citeauthoryear{Peebles} {1971}]{Peebles71} Peebles P. J.~E., Physical Cosmology.
Princeton University Press, 1971

\bibitem[\protect\citeauthoryear{Peebles} {1993}]{Peebles93} Peebles P. J.~E., Principles of Physical
Cosmology. Princeton University Press, 1993

\bibitem[\protect\citeauthoryear{Percival et al.} {2004}]{Percival04} Percival W. J., et al.,
2004, MNRAS, 353, 1201

\bibitem[\protect\citeauthoryear{Pike \& Hudson} {2005}]{Pike05} Pike R. W., Hudson M.
J., 2005, ApJ, 635, 11

\bibitem[\protect\citeauthoryear{Radburn-Smith et al.} {2004}]{Radburn-Smith04} Radburn-Smith
D.J., Lucey J.R., Hudon M.J., 2004, MNRAS, 355, 1378

\bibitem[\protect\citeauthoryear{Reid et al.} {2012}]{Reid12} Reid B. A., 2012
1203.6641 [arXiv:astro-ph]


\bibitem[\protect\citeauthoryear{Samushia et al.} {2012}]{Samushia12} Samushia L.,
Percival W. J., Raccanelli A., 2012, MNRAS, 420, 2102

\bibitem[\protect\citeauthoryear{Saunders et al.} {2000}]{Saunders00} Saunders W., et al.,
2000, MNRAS, 317, 55

%

\bibitem[\protect\citeauthoryear{Scaramella et al.} {1989}]{Scaramella89} Scaramella R.,
Baiesi-Pillastrini G., Chincarini G., Vettolani G., Zamorani G.,
1989, Nature, 338 652

\bibitem[\protect\citeauthoryear{Schlegel et al.}{2011}]{bigboss} Schlegel D., et al., 2011, arXiv:1106.1706

\bibitem[\protect\citeauthoryear{Scoccimarro et al.} {2001}]{Scoccimarro01} Scoccimarro R.,
Feldman H. A., Fry J. N., Frieman J. A., 2001, ApJ, 546, 652

\bibitem[\protect\citeauthoryear{Scott \& Smoot}{2010}]{Scott10}
Scott D., Smoot G., 2010 Review of Particle Physics, 1005.0555
[arXiv:astro-ph]

\bibitem[\protect\citeauthoryear{Skrutskie et al.}{2006}]{Skrutskie06} Skrutskie M. F., et al., 2006, AJ, 131, 1163


\bibitem[\protect\citeauthoryear{Square Kilometre Array} {}]{ska}
Square Kilometre Array: http://www.skatelescope.org

\bibitem[\protect\citeauthoryear{Springob et al.} {2007}]{Springob07} Springob C. M.,
Masters K. L., Haynes M. P., Giovanelli R., Marinoni C., 2007,
ApJS, 172, 599

\bibitem[\protect\citeauthoryear{Song \& Percival} {2009}]{Song09} Song Y.-S., Percival W. J.,
2009, JCAP, 10, 4

\bibitem[\protect\citeauthoryear{Strauss and Willick} {1995}]{Strauss95} Strauss M. A.,
Willick J. A., 1995, Phys. Rep., 261, 271

%

\bibitem[\protect\citeauthoryear{Tegmark et al.} {2006}]{Tegmark06} Tegmark M., et al.,
2006, Phys. Rev. D, 74, 123507

\bibitem[\protect\citeauthoryear{Tojeiro et al.} {2012}]{Tojeiro12} Tojeiro R., et al., 2012
MNRAS, tmp 3331T


\bibitem[\protect\citeauthoryear{Tonry et al.} {2000}]{Tonry00} Tonry J L.,
Blakeslee J. P., Ajhar E. A., Dressler A., 2000, ApJ, 530, 625

\bibitem[\protect\citeauthoryear{Tonry et al.} {2003}]{Tonry03} Tonry J. L. et al., 2003, ApJ, 594, 1

\bibitem[\protect\citeauthoryear{Turnbull et al.} {2012}]{Turnbull12} Turnbull S. J., Hudson
M. J., Feldman H. A., Hicken M., Kirshner R. P., Watkins R., 2012,
MNRAS, 420, 447

\bibitem[\protect\citeauthoryear{Verde et al.} {2002}]{Verde02} Verde L. et al., 2002, MNRAS, 335, 432

\bibitem[\protect\citeauthoryear{Watkins et al.} {2009}]{Watkins09} Watkins R., Feldman H. A., Hudson
M. J., 2009, MNRAS, 392, 743

\bibitem[\protect\citeauthoryear{Wenger et al.} {2003}]{Wenger03} Wenger G. et al., 2003, Astron. J,
126, 2268

\bibitem[\protect\citeauthoryear{Willick et al.} {1997}]{Willick97} Willick J. A., Strauss M. A., Dekel A.,
Kolatt T., 1997, ApJ, 486, 629

\bibitem[\protect\citeauthoryear{Willick et al.} {1998}]{Willick98} Willick J. A., Strauss M. A., 1998, ApJ,
507, 46

\bibitem[\protect\citeauthoryear{Willick} {1999} ] {Willick99} Willick J. A., 1999, ApJ,
522, 647

\bibitem[\protect\citeauthoryear{Yahil et al.} {1991}] {Yahil91} Yahil A., Strauss M.A., Davis M., Huchra J.P., 1991, ApJ, 372, 380

\bibitem[\protect\citeauthoryear{Zaroubi et al.} {2002}]{Zaroubi02} Zaroubi S., Branchini E., Hoffman Y., da Costa
L.N., 2002, MNRAS, 336, 1234






\end{thebibliography}

\end{document}